# Chemically Induced Transformation of CVD-Grown Bilayer Graphene into Single Layer Diamond

**Authors:** Pavel V. Bakharev[1*], Ming Huang[1,2], Manav Saxena[1], Suk Woo Lee[2], Se Hun Joo[3], Sung O Park[3], Jichen Dong[1], Dulce Camacho-Mojica[1], Sunghwan Jin[1], Youngwoo Kwon[1], Mandakini Biswal[1], Feng Ding[1], Sang Kyu Kwak[1,3], Zonghoon Lee[1,2] & Rodney S. Ruoff[1,2,4*]

**Affiliations:**

[1]Center for Multidimensional Carbon Materials (CMCM), Institute for Basic Science (IBS), Ulsan 44919, Republic of Korea.

[2]School of Materials Science and Engineering, Ulsan National Institute of Science and Technology (UNIST), Ulsan 44919, Republic of Korea.

[3]School of Energy and Chemical Engineering, Ulsan National Institute of Science and Technology (UNIST), Ulsan 44919, Republic of Korea

[4]Department of Chemistry, Ulsan National Institute of Science and Technology (UNIST), Ulsan 44919, Republic of Korea.

*Correspondence to: rsruoff@ibs.re.kr, ruofflab@gmail.com (RSR); bakharevpavel@gmail.com (PVB)

**Abstract**: Notwithstanding numerous density functional studies on the chemically induced transformation of multilayer graphene into a diamond-like film, a comprehensive convincing experimental proof of such a conversion is still lacking. We show that the fluorination of graphene sheets in Bernal (AB)-stacked bilayer graphene (AB-BLG) grown by chemical vapor deposition on a single crystal CuNi(111) surface triggers the formation of interlayer carbon-carbon bonds, resulting in a fluorinated diamond monolayer ('F-diamane'). Induced by fluorine chemisorption, the phase transition from AB-BLG to single layer diamond was studied and verified by X-ray photoelectron, ultraviolet photoelectron, Raman, UV-Vis, electron energy loss spectroscopies, transmission electron microscopy, and DFT calculations.

Graphene and diamond are two well-known carbon allotropes with $sp^2$ and $sp^3$ bonding hybridization, respectively, and are characterized by outstanding physical and chemical properties. Graphene is a single-atom-thick network of carbon atoms arranged in a hexagonal honeycomb lattice that has aroused a great deal of interest thanks to its high mechanical strength, high thermal and electrical conductivities, elasticity, etc. Diamond has an exceptionally high thermal conductivity, mechanical hardness and stiffness. There are a number of theoretical studies on the conversion of multilayer graphene into ultrathin 'diamond' films (these are so thin, just a few layers thick, that they have been given a new name: *diamane*) by attaching fluorine or hydrogen atoms, or hydroxyl groups, to the outer surface of graphene films (*1-5*).

There are two recent reports of generating interlayer bonding between graphene layers but only during application of very high pressure. Release of the pressure causes the return to the bilayer graphene, as described in both reports. One group reported formation of an unstable interlayer bonded material (stable at high pressure only) in a diamond anvil cell as per changes in the Raman spectrum (*6*). The other group reported local formation of interlayer bonding as a result of the high pressure applied by a silicone probe and a diamond indenter in an atomic force microscope (*7*); this region converted back to bilayer graphene when the high pressure was removed.

Our motivation was to make diamane over large area and that is stable at 1 atmosphere pressure. Our strategy included noting that graphene functionalization with F atoms has several advantages over hydrogenation—in our first attempts to make diamane. First, according to the Pauling scale, the electronegativities of carbon, hydrogen and fluorine have values of 2.55, 2.20 and 3.98, respectively. Covalent C–F bonds are strongly polarized toward the F atom. Therefore, a fluorinated graphene structure can be quantitatively characterized using X-ray photoelectron spectroscopy (XPS) due to a strong binding energy shift of the C–F peak towards higher binding energies relative to the C–C ($sp^2/sp^3$) peaks in the C1s spectrum. This allows us to (accurately) obtain the configuration and the C/F stoichiometry of the fluorinated structure by XPS. In the case of hydrogenated graphene, it is impossible to quantitatively distinguish between $C(sp^3)$–H and $C(sp^3)$–$C(sp^3)$ bonds in the C1s XPS spectrum (*8-10*). In contrast to hydrogen, fluorine can be "directly" identified by XPS, EDX (energy dispersive X-ray), and EELS (electron energy loss spectroscopy).

Second, common techniques of graphene hydrogenation/hydroxylation by "clean" gas-phase reaction require the use of "hot" hydrogen atoms or hydroxyl radicals that can be produced using hot filament (*8,9*), high-pressure (*6,11*), or plasma (*12,13*) methods, which (at least at this early stage of diamane research) make it difficult to control the H/OH coverage and the stoichiometry as well as the degree of induced defects. Fluorination can be performed under moderate conditions (near-room temperatures and at low/atmospheric pressure) e.g. by using xenon difluoride ($XeF_2$) vapor as a source of fluorine (*14-16*).

Here we provide the first experimental evidence that carbon-carbon interlayer bonds are formed by the fluorination of bilayer graphene (BLG), and that this results in the formation of a thin film of fluorinated diamond (F-diamane). In our study we used XPS, UPS (ultraviolet photoelectron spectroscopy), Raman spectroscopy, UV-Vis-NIR, EELS, transmission electron microscopy (TEM), scanning TEM (STEM), and density functional theory (DFT) to obtain structural information about F-diamane and to elucidate its formation mechanism.

In order to fluorinate graphene in a controllable and reproducible way, we used a home-built fluorination system that enabled us to control the partial pressure of the $XeF_2$ vapor and the temperature in the reaction chamber. By changing the fluorination conditions, such as temperature, $XeF_2$ partial pressure, and exposure time, we were able to obtain fluorinated graphene structures having different C/F ratios. (Next we provide a rather detailed description of our XPS studies that prove synthesis of F-diamane; we note that the reader can also immediately see a cross-section of F-diamane in Figures 3 and S9 obtained by HR-TEM and STEM, respectively).

XPS was used to quantitatively characterize the fluorine coverage and to probe the nature of the chemical bonds of the graphene samples before and after fluorination. High-resolution core levels and survey spectra were acquired for the as-grown graphene and for fluorinated films. Figure S1 shows the survey and core levels (C1s, Cu2p, and Ni2p) of as-grown bilayer graphene (BLG) on a single crystal CuNi (111) alloy foil. The C1s spectrum of the BLG shows a single narrow asymmetric peak at 284.2 eV with a 0.58 eV full width at half maximum (FWHM), which is a signature of $sp^2$ hybridized carbon network. The high-resolution Cu2p and Ni2p spectra of the CuNi alloy substrate before fluorination correspond to metallic Cu and Ni (*17,18*).

Graphene films were fluorinated at 65 ºC under 50-60 Torr vapor pressure of $XeF_2$. Angle-resolved XPS (AR-XPS) was used to qualitatively study the distribution of chemical bonds over the layers by comparing relative intensities of the corresponding peaks obtained at 0º and 50º grazing emission angles between the surface normal and the direction toward the detector. The measurements taken at a 50º emission angle are more surface sensitive than those at 0º because of the difference in the photoelectron escape depths (*19*). After fluorination of the BLG on the CuNi (111) surface, XPS revealed three typical structures of fluorinated BLGs and we shall refer to them as Sample A (the final configuration obtained after >12 hour fluorination), Sample B (~6 hour fluorination) and Sample C (2-3 hour fluorination) (we note that there is a purge/pump step in our process for exposing to XeF$_2$ that takes ~1h). The high-resolution C1s and F1s spectra of Samples A, B, C are shown in Figs. 1(a-c) and Figs. S2(a-c), respectively. The carbon atoms directly bonded to fluorine atoms contribute to the C-F peak at 288.1 eV while the peaks in the binding energy range from 284 eV to 286 eV correspond to carbon-carbon (C-C) bonds (*20-26*). The features at 291-293 eV are attributed to C-F$_2$ and C-F$_3$ groups that are formed at the edges and on the structural defects in the graphene sheets (*22*). The peaks at 285.5-285.6 eV (Samples B and C) and at 287.2-287.4 eV correspond to "bare" (not directly bonded to fluorine) carbon atoms adjacent to C-F and C-F$_{2,3}$ groups, respectively (*21-25*). Strong shifts of these peaks (referred to as C-CF and C-CF$_{2,3}$ in Fig. 1) to higher binding energies relative to C=C double bonds can be explained by the perturbation of the electronic structure of the "bare" C in the vicinity of covalent C-F and C-F$_{2,3}$ bonds due to the high electronegativity of fluorine. As can be seen in Fig. 1, prolonged fluorination (>12 hrs) of BLG on a single crystal CuNi (111) surface under the conditions mentioned above results in the formation of the fluorinated structure, Sample A, with an essentially perfect C$_2$F stoichiometry (see Table S1 for details on the estimation of fluorine content in Samples A, B, and C).

There are two possible structures for the C$_2$F film. The first structure (Fig. S3(a)) is characterized as a single diamond layer (F-diamane) with sp$^3$-hybridized carbon atoms that participate in either interlayer C-C or surface C-F bonds (*20*). In the second structure (Fig. S3(b)) C-C interlayer linkages are not established and fluorine atoms are bound to both sides of each graphene sheet, forming –C=C–C=C– chains (*23-25*). Based on our DFT study of various configurations with no interlayer bonds (Fig. S4), the interlayer spacing of these structures ranges from 0.38 nm to 0.49 nm. In Fig. S3(b) we show a model of the most energetically favorable $C_2F$ arrangement with no interlayer linkages. Following our spectroscopic and microscopic studies, and theoretical calculations, we have proved the formation of the former diamond-like structure (F-diamane), see below. F-diamane is the most energetically favorable among all the $C_2F$ configurations (Fig. S4).

By comparing the intensity ratios, $I_{sp2}/I_{C-CF}$, of the peak corresponding to $sp^2$ carbon (see Fig. 1) to the nearest peak (C-CF) associated with the presence of C-F bonds in the C1s spectra acquired at 0º and 50º emission angles, we can *qualitatively* analyze the distribution of C-F bonds over the layers. Thus, if the ratio, $R$, of $I_{sp2}/I_{C-CF}(0º)$ measured at 0º emission angle to $I_{sp2}/I_{C-CF}(50º)$ obtained at 50º emission angle is close to unity than it indicates an even distribution of C-F covalent bonds over the layers; if $R > 1$ then the top layer of fluorinated BLG has a higher F content than the bottom layer and if $R < 1$ then the top layer is fluorinated less than the bottom layer. According to the AR-XPS analysis shown in Figs. 1(a,b,c), Sample A of an essentially $C_2F$ configuration has the ratio $R$ equal to 0.97±0.05 and hence has a more even distribution of C-F bonds over the layers, while in the cases of Samples B ($R$=1.14±0.05) and C ($R$=1.37±0.05) the bottom layer has a lower fluorine content than the top layer. Also, in the binding energy range of 284–286 eV, we see 3 distinct peaks for Sample B while there are only 2 well-pronounced peaks in that range for Samples A and C. The peaks at 284.9 eV–285.3 eV are ascribed to the C-C $sp^3$ component and can be attributed to the formation of interlayer bonds. In the binding energy range 284.9 eV–285.6 eV two peaks, namely C-CF at 285.6 eV and C-C $sp^3$ at 284.9 eV, have been resolved for Sample B, whereas only one peak at 285.2 eV–285.3 eV has been detected for Sample A and one peak at 285.5 eV has been observed for Sample C. We assume that the component at 284.9 eV in Sample B originates primarily from the carbon atoms in the bottom layer that have formed interlayer bonds with the carbon atoms in the top layer. Our assumption is based on the fact that since the bottom layer in Sample B is less fluorinated than the top layer, then the C atoms in the bottom layer which form the interlayer bonds can generate a "pure" sp$^3$ C-C signal at 284.9 eV, which is not strongly affected by the presence of carbon-fluorine bonds. At the same time, based on AR-XPS, the signals at 255.6 eV corresponding to the C-CF components in Samples B and C originate primarily from the top layer. In contrast, in Sample A it is impossible to distinguish between the C-CF and C-C sp$^3$ components due to an essentially perfect C$_2$F stoichiometry and an even distribution of C-F bonds over the layers. Thus, every second C atom in Sample A, which may form an interlayer bond, is also adjacent to C atoms bonded to fluorine so that it is impossible to resolve "pure" C-C sp$^3$ and C-CF signals. That explains the emergence of a single C-CF/sp$^3$ peak in the binding energy range 284.9 eV–285.6 eV. In Sample C with a relatively low fluorine content (15-16 at.%), the signal corresponding to the C-C interlayer bonds has not been evidently detected because of a lack of C-F covalent bonds which can stabilize the structure with interlayer linkages. A time dependent XPS analysis of fluorinated Samples A, B and C was conducted and it was found that the structures of Samples A and C are considerably more stable than the structure of Sample B (Fig. S5).

The deconvolution of the F1s spectra (Fig. S2) revealed signals corresponding to so-called 'semi-ionic/semi-covalent' F-C bonds (686.7–686.9 eV) (23-26), ionic compounds (684.4–684.7 eV) (21), 'physisorbed/entrapped' F (681.3–681.6 eV) (27), and covalent F-C bonds (688.0–688.2 eV) (20,21,27). The AR-XPS showed that (i) the ionic signal is attributed to the fluorine bonded to metal atoms, and (ii) the F1s signal at 681.3–681.6 eV originate primarily not from the top layer and the surface but from the bottom layer and the graphene-metal or graphene-metal fluoride interface, respectively. These conclusions are based on the attenuation of the ionic and the 'physisorbed/entrapped' peaks at the 50º emission angle relative to the main F-C component at 686.7–686.9 eV in the F1s spectra and relative to the C1s signal. The AR-XPS analyses of C1s and F1s spectra allow us to estimate the average thickness of the fluorinated BLG "over-layer" (see Table S2 for details). The fact that the signal 681.3–681.6 eV is from the interface and not from fluorine intercalated between the graphene sheets has been tested in our analysis of fluorinated graphene film "suspended" on a TEM grid (as discussed below). The positions of the main C-F peaks at 686.7–686.9 eV and at 287.9–288.2 eV in the F1s and C1s spectra, respectively, agree well with the corresponding values reported for 'semi-covalent/semi-ionic' C-F bonds in fluorinated graphite/graphene structures (*23-26*). The covalent signal at 688.0–688.2 eV is assigned to C-$F_{2,3}$ bonds (*27*).

In order to characterize the nature of the C-F bonds in fluorinated BLG, we analyzed binding energy gaps between primary C-F peaks in the C1s and F1s spectra. Values ranging from 398.6 to 398.8 eV were obtained for all of our fluorinated BLG samples, consistent with covalent C-F bonding (*26*).

The ionic signal at 684.4–684.7 eV in Fig. S2 and the high binding energy satellite peaks in the Cu2p and Ni2p spectra shown in Fig. S6 indicate the formation of metal fluoride compounds, $CuF_2$ and $NiF_2$. This means that the graphene membrane is permeable to fluorine because of defects in the as-grown CVD graphene and (perhaps also) its further damage during fluorination. The emergence of C-$F_2$, C-$F_3$ and possibly C=O groups also confirms the presence of edges and defects in the graphene sheets.

The permeation of fluorine through the graphene membrane and further fluorination of the metal surface indicate the presence of fluorine at the BLG-metal interface that may lead to the fluorination of BLG from both sides, namely from the open surface of the top layer and from the bottom layer at the metal fluoride-graphene interface, and result in the formation of a double-layer of $C_2F$ stoichiometry stabilized with C-C interlayer linkages (Fig. S2(a)) as suggested by the XPS. The formation of $C_2F$ configuration in Sample A (>12 hour fluorination) through the intermediate states of Samples C (2-3 hour fluorination) and B (5-6 hour fluorination) suggests the mechanism of F-diamane formation on the metal surface, namely, the sequential fluorination of the top and bottom layers (top then bottom) in a relatively defective CVD-BLG. Defects in as-grown graphene may play a critical role in the emergence of the interlayer C-C bonds, nucleation sites, as a result of uneven fluorination of the BLG film as shown by our XPS analysis of Samples B and C. In that case, F atoms at the interface form covalent bonds with the bottom layer to stabilize the local structure with interlayer C-C linkages creating diamane nuclei. Further fluorination induces lateral propagation of C-C interlayer bonds resulting in formation of the fluorinated diamond monolayer of an essentially perfect $C_2F$ stoichiometry (Sample A). It is noteworthy that no increase in the C-F covalent bonds content was observed after fluorination for more than 12 hours, confirming complete conversion of BLG into a $sp^3$-hybridized structure.

Figure S7 shows the calculated energies of pristine BLG with the unit cell containing 16 C atoms ($C_{16}$) and fluorinated BLG of different stoichiometry, namely of $C_{16}F_2$ with two F atoms on the top layer, $C_{16}F_3$ with an interlayer C-C bond stabilized with two F atoms on the top layer and one F atom on the bottom (diamane nucleus), $C_{16}F_4$ with a C-C interlayer linkage and two F atoms bonded to each layer, and $C_{16}F_8$ (F-diamane). These configurations describe the evolution of BLG to F-diamane during fluorination. The fact that the energy of $C_{16}F_3$ (with an interlayer C-C bond) is higher than the energy of $C_{16}F_2$ (with no interlayer linkages) indicates the presence of a diamane nucleation barrier. If the barrier is high, the interlayer bond formation may start from a relatively defective CVD-BLG on the metal surface. A further increase in the fluorine content in the system ($C_{16}F_x$ and x>3) induces the formation of more interlayer C-C bonds and leads to decrease in energy. Once the bilayer structure "zips" all the way through, the phase transition is complete and highly stable F-diamane, as indicated by its very low energy, is formed.

The Raman spectra of as-grown BLG on the CuNi (111) surface, as well as the Raman signatures of Samples A, B and C are shown in Fig. 2. The position, the FWHM, the shape of the 2D band and the ratio of the G to the 2D peak intensities obtained for the as-grown BLG transferred onto a $SiO_2$/Si wafer are typical of those of AB-stacked (Bernal) BLG (AB-BLG) (Fig. S8) (*28-30*). The Raman spectra of Sample C is shown in Fig. 2(b). The emergence of a prominent D band at 1360 cm$^{-1}$ indicates a strong modification of CVD-grown BLG on the CuNi(111) due to exposure to $XeF_2$. As fluorination time increased to ~6 hours (Fig. 2(c)), the D peak increased in intensity and shifted below 1350 cm$^{-1}$ while the 2D/G' graphene Raman decreased in intensity. In some regions in Sample B, the Raman signatures were almost completely suppressed indicating a high transparency of the fluorinated BLG structure. The Raman characterization of Sample A with 32-33 at.% of fluorine content showed a uniform structure over a large area with suppressed Raman signatures; this structure resembles that of fluorographene (*14,15*). It is worth mentioning that in fluorographene every carbon atom is covalently bound to a fluorine atom so that all carbon atoms are sp$^3$ hybridized. In the case of *partially* fluorinated BLG (31-32 at.% of fluorine), the disappearance of the Raman signatures of sp$^2$ coordinated carbon atoms can be rationalized by the formation of a double layer $C_2F$ structure with interlayer C-C linkages, which is schematically shown in Fig. S3(a). Relatively weak D and G Raman signals in Sample A were observed in the isolated regions (domains) of a nearly hexagonal shape. Figure S9 shows Raman maps for as-grown BLG on CuNi (111) as well as for Samples A, B, and C. Sample A showed the quenched Raman signal except for the domains with presumably trilayer graphene. We found hexagonal-shaped ABA-stacked trilayer graphene 'patches' in as-grown graphene film transferred onto SiO$_2$/Si wafer (with 300 nm thick SiO$_2$ surface layer) as shown in Fig. S10. These isolated regions in Sample A can be distinguished under the optical microscope on the metal fluoride surface (Figs. S9 and S11) due to the difference in transparency of continuous C$_2$F BLG (F-diamane) and the regions with presumably fluorinated ABA-stacked trilayer graphene domains (Figs. S9 and S11). At the same time these regions are not visible either in the as-grown BLG on the CuNi (111) surface, or in Samples B and C with lower fluorine contents. It is noteworthy that ABA…-stacked multilayer graphene, including trilayer and thicker graphene films, cannot be converted into "thick" diamanes because of the unfavorable Bernal (*ABA* …) layer stacking sequence and thus the carbon atoms in every 3$^{rd}$ layer are in "wrong" positions to form interlayer C-C linkages with the 2$^{nd}$ layer. The layer stacking that would favor diamane formation for 3 or more layers is rhombohedral (*ABC*...). Hence, our optical microscope images and the Raman characterizations show the large area 'transparent' regions in Sample A (of essentially C$_2$F stoichiometry), which indirectly proves the formation of *large area* F-diamane.

Ultraviolet photoelectron spectroscopy (UPS) was carried out to characterize the valence band states in as-grown BLG as well as in fluorinated films with a different fluorine content (Samples A and B). The corresponding valence band spectra of as grown BLG, Samples A and B were collected using monochromatized He I (21.2 eV) radiation and are shown in Fig. S12. The UPS spectrum of CVD-BLG contains a small peak A at ~0.8 eV, peaks B at 3 eV, C at 5.8 eV, D at 10.4 eV and E at 13.7 eV below the Fermi level; these peaks correspond to the states around the K point (A), the π band formed by 2p$_z$ orbitals (B), the crossing of 2p-π and 2p-σ bands (C), the p-like σ band (D) and the high density of states (DOS) in the σ band formed by hybridized 2s, 2p$_x$, 2p$_y$ orbitals (E) (*30,31*). A signal below 15 eV (band F) is attributed to the s-like DOS. After fluorination we observed the emergence of the states at the range 10–11.5 eV below the valence band maxima ascribed to the fluorine 2p-like states (*32*). The UPS spectrum of Sample A had two sharp peaks at 11.2 eV (fluorine 2p-like states) and 12.6 eV below the valence band maximum. The peak at 12.6 eV can be assigned to the diamond-like carbon 2s-2p hybridized orbitals (*33*). The work function of pristine BLG and the ionization potentials of the fluorinated samples were calculated by subtracting the width of photoelectron spectrum from the photon energy. Thus, the work function of as-grown BLG was found to be 4.4 eV. The ionization potentials of 6 eV and 8 eV were obtained for Samples B and A, respectively. The significant decrease in the spectral width after fluorination can be rationalized by the polarity of carbon-fluorine covalent bonds. The ionization potential of 8 eV measured for Sample A agrees well with the value of 8.71 eV estimated for F-diamane (for details see the supplementary information Fig.S13). It is worth noting that the ionization potential of ~8 eV was reported for the fluorine-terminated diamond surface (*34*).

In order to explore the atomic arrangements in fluorinated BLG on the single crystal CuNi (111) surface, TEM was used. Experimental cross-sectional high resolution TEM (HR-TEM), scanning TEM (STEM) and simulated HR-TEM images are shown in Fig. 3 and Fig. S9. The two-layer structure of the pristine graphene with a 3.24–3.41 Å interlayer separation was significantly altered upon exposure to $XeF_2$. After fluorination, our TEM/STEM study revealed a highly-ordered structure on the CuNi(111) surface with characteristic interlayer/interatomic distances shown in the right image in Fig. 3 and in Fig.S14(b). To elucidate the ordered arrangements of F and C atoms in the fluorinated BLG, simulated HR-TEM images were constructed for the DFT-optimized $C_2F$ configurations with and without interlayer linkages as shown in Fig. 3 (the lower images) and in Fig.S14. Since neither an increase in interlayer spacing nor the characteristic atomic arrangements with F atoms alternately bound to each graphene layer from both sides were detected after fluorination, the emergence of the $C_2F$ configurations without interlayer C-C bonds, including the most energetically favorable structure shown in Fig.S15, can be ruled out. The simulated cross-sectional TEM images for (110) and (100) planes of DFT optimized F-diamane are given in Fig. 3. It can be inferred from Figs. 3, S14(b) and S15 that the atomic arrangements and the interlayer/interatomic separations in the TEM/STEM images obtained after the fluorination of BLG on a CuNi (111) surface match well those of the simulated F-diamane structure.

BLG fluorinated on a single crystal CuNi (111) surface was then transferred onto Au TEM grids (we refer to this transferred structure as F-BLG) by using the electrochemical bubbling delamination method. We used two types of the grids, namely, Au TEM grids covered with Quantifoil (2 μm hole size) and bare Au grids. The former, due to the good adhesion between graphene and a holey carbon support film, enables us to obtain a high quality large area suspended BLG for HR-TEM, EELS, Raman and optical studies. The latter were used for the XPS characterization to determine the C/F ratio in suspended BLG without the "parasitic" signal from the polymer support film.

In order to recover the fluorine content after the transfer process (Fig. S16), the F-BLG on the TEM grid was exposed to XeF$_2$ vapor. According to the XPS and Raman analyses (Figs. 4(a,b)), the C$_2$F structure with quenched Raman graphene modes can be obtained after a 6 hour fluorination of this F-BLG at 45°C. A comparatively weak F1s ionic signal detected in the suspended sample is ascribed to the gold fluoride compound, the formation of which upon exposure of the Au TEM grid to XeF$_2$ has also been verified by the emergence of prominent satellite peaks in the Au4f HR-spectrum (Fig. S17). The ratio, $I_{phys}/I_{C-F}$, of the integrated intensity of the F1s peak at 681.7 eV to the integrated intensity of the primary F-C peak at 686.7 eV in Fig. 4(a) (0.02) is one fifth of that obtained for a graphene film fluorinated on the metal surface (0.1). This confirms that the primary source of the signal at 681.5 eV in the F1s spectra shown in Fig. S2 is fluorine trapped between the bottom graphene layer and the metal fluoride surface.

UV-Vis-NIR absorption was measured for the films suspended on the Au TEM grid. The light beam was focused with a 52x objective in order to sample an area 1.5 μm–1.6 μm in diameter, thus within the holes in the carbon support film. Fig. 4(c) shows representative absorption spectra of pristine and fluorinated BLG. The asymmetric absorption peak at 4.6 eV with ~6% absorption in the near-IR range obtained for the pristine sample is in a good agreement with the data reported for CVD-BLG (*30*). For a fluorinated graphene film, significantly greater transparency in the visible range is attributed to the opening of a wide optical gap. According to the approach described in (*14,35*) for 2D electronic systems, the optical gap can be estimated by using a linear approximation, as shown in Fig. 4(c) by the dashed line; the optical gap of fluorinated BLG was thus found to be in the range 3.3–3.4 eV. The experimental optical gap value of 3.3–3.4 eV is consistent with the calculated optical gap of 2.87 eV obtained for F-diamane (in comparison with the electronic band gap of 1.31 eV of the most energetically favorable $C_2F$ structure with no interlayer bonds). The calculated optical gap value of 2.87 eV was obtained by introducing an F monovacancy defect per unit cell (as shown in Fig. S18); this monovacancy was found to induce a strong excitonic effect (see supplementary information for the details). The fact that the calculated value of the optical gap is lower than the experimental value can be rationalized by a relatively high concentration of F vacancies in our model of the '*defective*' F-diamane, namely 12.5%, which corresponds to the $C_2F_{0.88}$ configuration.

Figs. 5(a,b) show the HR-TEM images and the selected area electron diffraction (SAED) patterns obtained for the fluorinated and non-fluorinated BLG, respectively. The number of layers in the graphene films was verified by *in situ* gradual, layer by layer, etching of the graphene sheets with an electron beam at an acceleration voltage of 80 kV (Fig. S19) (*36*). The SAED patterns

were acquired from circular regions of ~0.7 μm diameter. The TEM data provided in Fig. 5 indicate a perfect in-plane hexagonal crystalline order in fluorinated and pristine BLGs. Detailed information on the atomic arrangements was gained by comparing the experimental TEM data with the simulated HR-TEM images of the DFT-optimized F-diamane (Fig. 5a), AB-BLG (Fig. 5b), $C_2F$ configuration without interlayer bonds (Fig. S20(a)) and ABA-stacked trilayer graphene (TLG) (Fig. S20(b)). We examined the HR-TEM images by taking line profiles from the experimental and the simulated micrographs, as indicated with the red and light blue lines and the corresponding insets in Fig. 5. In the micrographs of the fluorinated BLG and in the simulated TEM images of the DFT-optimized F-diamane the contrast/brightness of individual dots is uniform. The corresponding line profiles measured for the fluorinated film are similar to those obtained for the simulated F-diamane. In contrast to the fluorinated film, in the pristine BLG case (Fig. 5(b)) six dots arranged in hexagonal rings appear to be brighter than the dots located at the center of these rings. We also analyzed the intensity distribution over the diffraction peaks in the SAED patterns as well as in the digital diffractograms acquired for the DFT-optimized structures. For the obtained fluorinated film and the simulated F-diamane, the intensities of the 1$^{st}$ order diffraction peaks are higher than those of the 2$^{nd}$ order with the experimental intensity ratio ($I_1/I_2$) values of 2.7–2.8. For both the BLG and the simulated AB-BLG, the 1$^{st}$ order diffraction peaks have lower intensities than the 2$^{nd}$ order ($I_1/I_2$=0.4–0.5), which indicates the AB (Bernal) stacking sequence in the double-layer graphene film (*28,29*). Hence, the micrographs and the diffraction patterns of the fluorinated and non-fluorinated BLGs are a good match to the corresponding simulated data obtained for F-diamane and AB-stacked BLG, respectively, as shown in Fig. 5. By decreasing the size of the parallel electron beam from 11.1 μm to 0.9 μm and hence by increasing the electron beam dose rate by more than 150 times, we observed the evolution of the SAED pattern of the fluorinated BLG from the characteristic diamane-like diffraction pattern to that of AB-BLG, as shown in Fig. S21. The recovery of the AB-BLG structure is rationalized by electron beam induced defluorination of the graphene film as reported by Withers *et al.* (*37*) and Martins *et al.* (*38*). Thus, electron beam irradiation of F-diamane can in the future be used as a nanopatterning technique to construct novel electronic devices by selectively restoring AB-BLG regions in an insulating ultrathin diamond-like structure. The SAED patterns collected at different locations across the fluorinated BLG film (each SAED pattern was acquired from circular region of ~0.7 μm diameter) indicate single crystallinity over a relatively large area (Fig. S22).

F-diamane was further verified by TEM-EELS characterization. The K-edge and low electron energy loss (EEL) spectra of the fluorinated and pristine AB-BLG are shown in Figs. 5(c,d). The K-edge spectral regions give important information about the bonding configuration ($sp^2/sp^3$ in carbon materials) and can be used for quantitative elemental/structural analysis while the low energy losses (<100 eV) correspond to plasmons and interband excitations. In the pristine sample, the K-edge spectrum of carbon atoms shows a clear $sp^2$ signal with sharp energy loss peaks at 285 eV (1s – 2p($\pi$*)) and 292 eV (1s – 2p($\sigma$*)). In the fluorinated BLG, the carbon K-edge spectrum is dominated by 1s – 2p($\sigma$*) features at 293 eV and 297 eV, whereas the 1s – 2p($\pi$*) energy loss is strongly suppressed indicating a lack of $\pi$* orbitals. The disappearance of $\pi$* anti-bonding states after fluorination of BLG was also confirmed by low electron energy loss spectroscopy at energies below 60 eV. At the same time the C/F ratio estimated from the representative K-edge spectral regions of carbon and fluorine (Inset in Fig. 5(c)) is ~2.2 ($C_2F_{0.9}$) that is commensurate with the $C_2F$ stoichiometry identified by XPS. The main peak in the F K-edge spectral region at 689 eV originates from the covalently bonded fluorine (*39*). Thus, the EELS data shown in Figs. 5(c,d) support the formation of a predominantly $sp^3$–hybridized carbon structure (essentially, F-diamane with close to $C_2F$ stoichiometry) by the partial fluorination of AB-BLG. For comparison in Fig. 5(c) we also provide the calculated K-edge and low EEL spectra of F-diamane.

The results of the spectroscopic (XPS, Raman, UPS, UV-Vis, EELS) and microscopic (TEM, STEM) studies reported here indicate that fluorine chemisorption on CVD-grown BLG can result in a fluorinated diamond monolayer (F-diamane). The obtained diamond-like film is an ultra-thin wide-band gap semiconductor, whose electronic properties are highly dependent on surface termination species and thus has potential for applications in nano-optics, nanoelectronics, and can serve as a promising platform for micro- and nano-electromechanical systems. Also, F-diamane can possibly be used as a seed layer for growth of high quality single-crystal diamond films by CVD methods under moderate conditions (pressure and temperature).

**Methods**

Preparation of CuNi alloys:

CuNi alloys were prepared by alloying single crystal Cu(111) foils with Ni (~20 at% content) using electroplating and high temperature (1050°C) annealing methods. Cu foil (20 μm, 99.9%, Nilaco Co., Japan) was heated at 1050 °C with 10 sccm Ar and 10 sccm $H_2$ at atmospheric pressure (1 atm) for 12 h to convert it to a Cu(111) foil. Ni layers were then plated on the Cu(111) foil in an electrolytic solution, which was prepared by dissolving 140 g of $NiSO_4 \cdot 6H_2O$, 4 g of $NiCl_2 \cdot 6H_2O$, 2 g of NaF and 15 g of $H_3BO_3$ in 500 mL of deionized water. The current density in all the plating experiments was 0.02 A cm$^{-2}$. After washing and drying, the Ni-plated Cu(111) foils were placed in a quartz furnace and heated at 1050 °C for 4−6 h in a gas flow of Ar (20 sccm) and $H_2$ (20 sccm) at atmospheric pressure.

Synthesis of bilayer graphene films

BLG was synthesized on single crystal CuNi(111) by LPCVD. Briefly, the LPCVD was performed at $2.5\times10^{-4}$ Torr pressure and 1075°C by flowing an Ar/H$_2$ (10/1 ratio) mixture and CH$_4$(g) for 2 hrs through the CVD chamber. Details of the preparation procedure will be reported in a separate paper.

Electrochemical bubbling delamination

Graphene grown by CVD on the single crystal CuNi(111) surface was spin-coated with a polymethyl methacrylate (PMMA) layer at 3000 rpm for 1 min to provide mechanical support for the transfer. The PMMA/graphene/Cu-Ni(111) alloy foil stack was then dipped into a NaOH aqueous solution (1M) to act as the cathode in an electrolysis cell with a constant current supply. The PMMA/graphene layer was detached from the Cu/Ni(111) foil after tens of seconds as a result of the formation of a large number of H$_2$ bubbles at the interface between the graphene and the Cu/Ni(111) foil. After cleaning with deionized water, the floating PMMA/graphene layer was transferred to the target substrate (TEM grid or a SiO$_2$/Si wafer). Finally, the sample was dried and the PMMA was removed with acetone.

X-ray photoemission spectroscopy (XPS)

High resolution and survey spectra were acquired using the ESCALAB 250Xi XPS system with monochromatic Al K$\alpha$ (E=1.487 keV) X-rays and a 0.5 m Rowland circle monochromator. The ultimate system energy resolution is ~0.40 eV. Angle resolved XPS (AR-XPS) measurements were conducted by tilting a sample stage and hence by changing the electron emission angle from 0º to 50º degree between the surface normal and the direction toward the detector.

Raman characterization

Micro Raman measurements were conducted using a confocal Witec spectrometer with a 488 nm laser in the backscattering mode. A 100x objective with a laser spot size of about 400 nm was used. In order to avoid significant heating and any other damage such as desorption of the adatoms from the graphene, the laser power was below 1 mW and the spectral resolution was ≈ 3 cm$^{-1}$.

Transmission electron microscopy (TEM)

TEM, scanning TEM (STEM) and electron energy loss spectroscopy (EELS) characterizations were performed using an aberration-corrected Titan cube G2 operated at 80kV. Selected area electron diffraction (SAED) patterns were acquired by using a ~0.7 μm aperture. High resolution TEM images were collected with a 0.2s exposure time promptly after taking the SAED patterns to avoid significant electron beam damage (or defluorination) and images were filtered by the Wiener filtering method. EEL spectra were measured by a GIF Quantum ER system and the TEM-EELS method with ~11 μm beam size and 5mm entrance aperture was used to minimize beam damage (or defluorination).

Optical characterization

The absorption spectra of the graphene films were acquired by using a CRAIC 20/20 PV UV-Vis-NIR microspectrophotometer. The measurements were conducted by focusing the light beam with a 52X objective in order to sample an area of 1.5 μm–1.6 μm in diameter within the holes in carbon support film.

**Acknowledgments:** We acknowledge support from the Institute for Basic Science (IBS-R019-D1).

**Author contributions:** RSR conceived of the experiment. PVB wrote the manuscript, designed and constructed the experimental setup, did experiments, characterizations, and data analyses. PVB and RSR revised the manuscript. MH prepared CuNi(111) alloys by electroplating and annealing, synthesized and characterized graphene films on CuNi(111) alloys, and did graphene transfer onto TEM grids and $SiO_2$/Si wafers; MB participated in making metal alloy foils and synthesis of graphene films. PVB and MH contributed equally to this work. MS conducted the experiments and characterizations (transfer of the samples, assisting with building experimental setups as well as with XPS and Raman characterizations). SJ and YK converted polycrystalline commercial Cu foils into single crystal Cu(111). SWL and ZL did TEM/STEM/EELS characterizations and TEM image simulations. SHJ, SOP and SKK did DFT optimization, and calculated formation energies as well as electronic and optical band gaps for various $C_2F$ configurations (Fig.S4, Fig.S13, Fig.S18). JD and FD did the DFT calculations to simulate the transformation of BLG into F-diamane (Fig.S7), and EELS simulations. DCM did DFT calculations, electronic band structure calculations, and the TEM/DP image simulations.

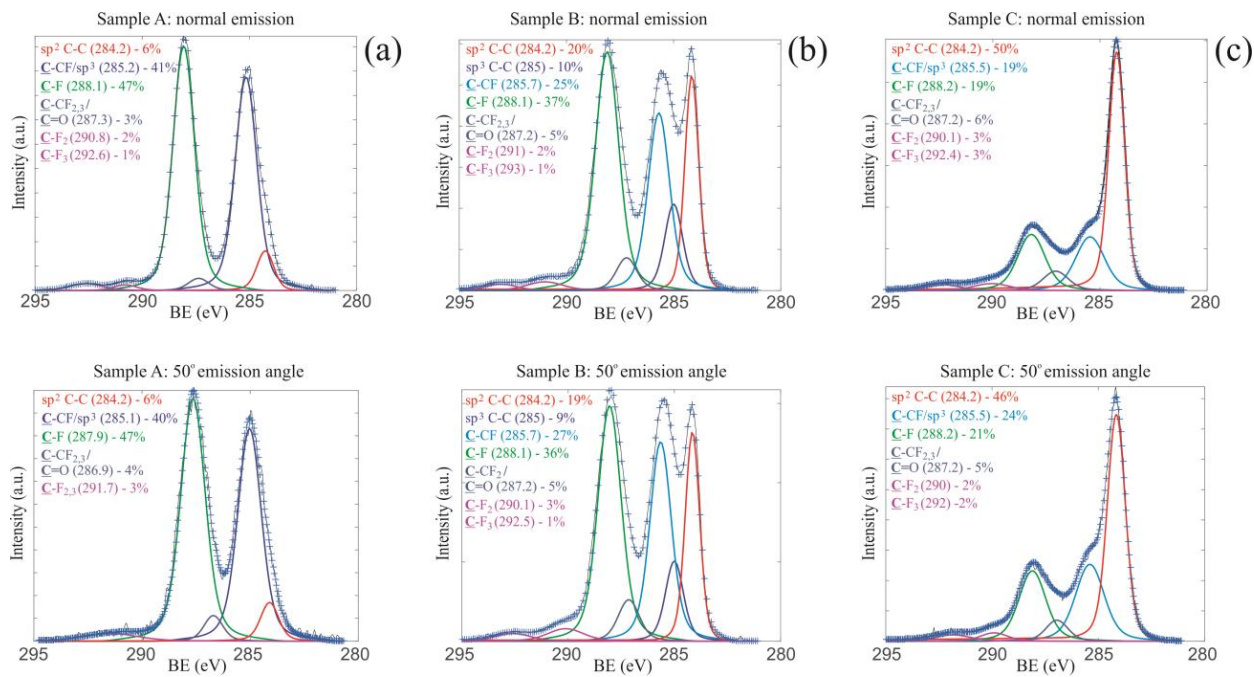

**Fig. 1. AR-XPS characterization of fluorinated BLG.** C1s spectra obtained at 0° and 50° grazing emission angles for (a) Sample A (>12 hour fluorination), (b) Sample B (~6 hour fluorination) and (c) Sample C (2–3 hour fluorination).

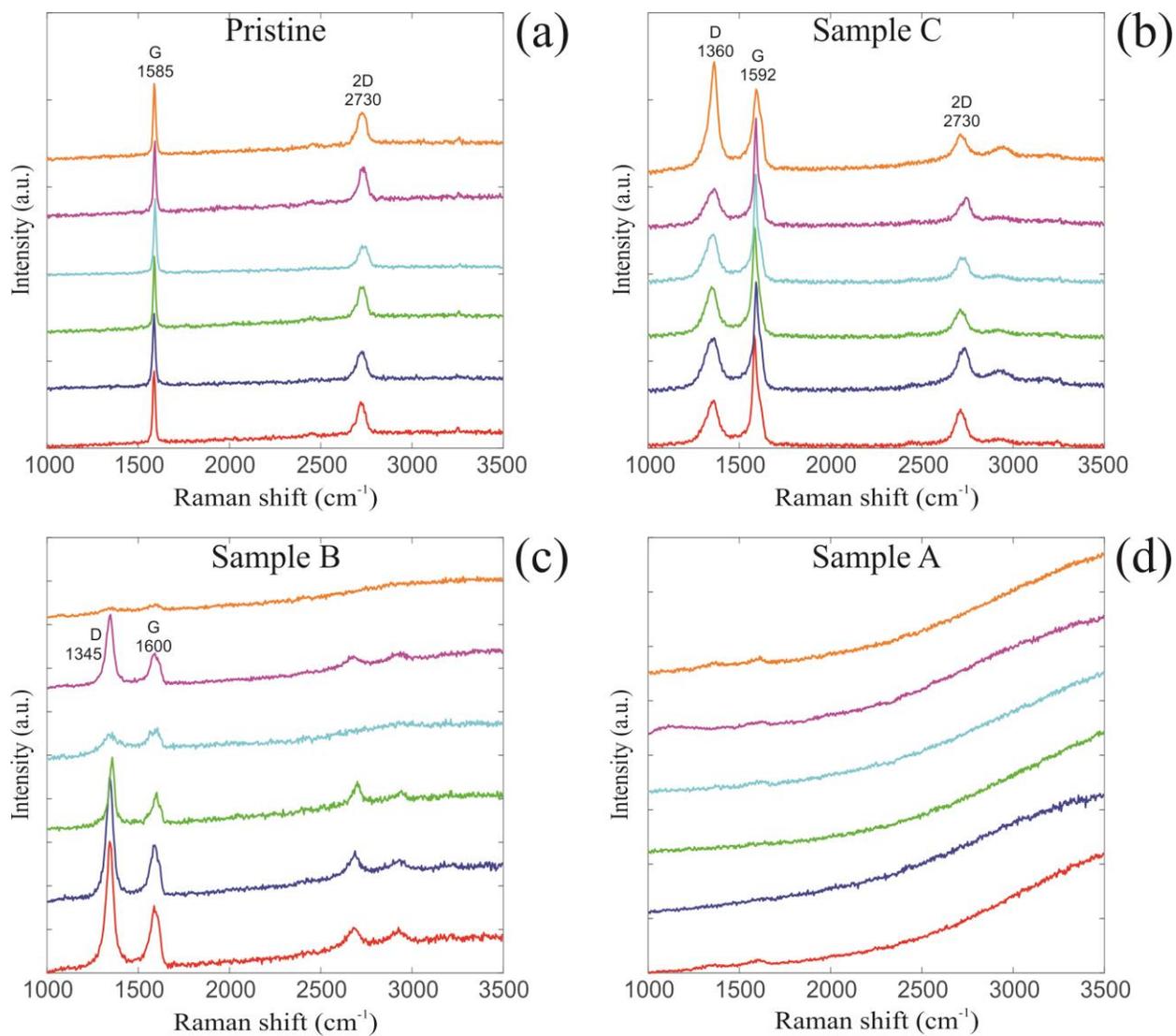

**Fig. 2. Raman characterization of fluorinated BLG on CuNi (111) surface.** Raman spectra at 6 randomly chosen positions by 488 nm excitation of (a) as-grown BLG, (b) Sample C (2–3 hour fluorination), (c) Sample B (~6 hours fluorination) and (d) Sample A (>12 hours fluorination)..

**Fig. 3. TEM study of fluorinated BLG on CuNi (111) surface.** High resolution cross-sectional transmission electron micrographs of as-grown (pristine) BLG (top left image) and Sample A (top right images). Simulated HR-TEM images of DFT-optimized F-diamane (bottom images).

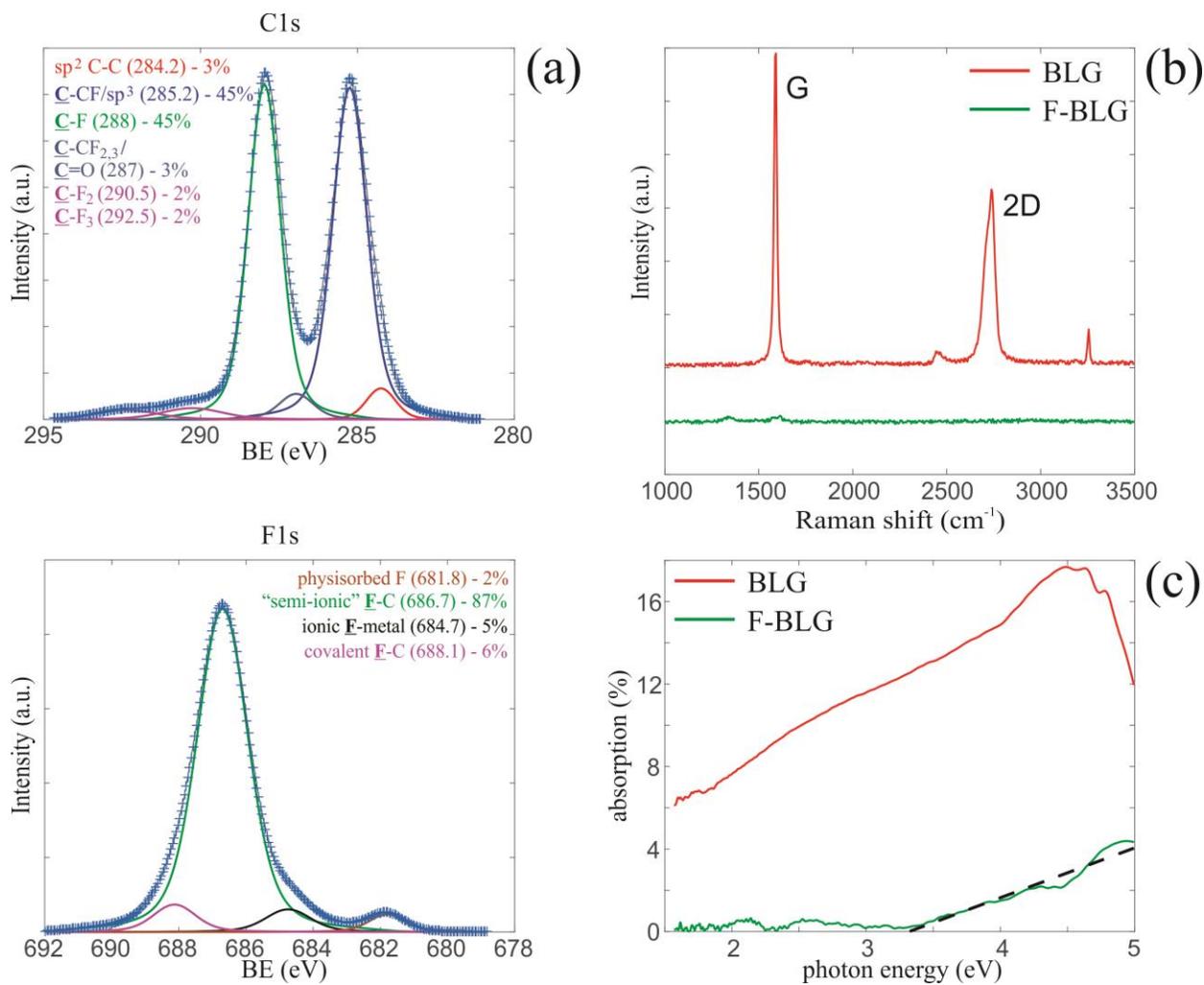

**Fig. 4. Spectroscopic analyses of fluorinated CVD BLG "suspended" on TEM grid.** (a) XPS C1s and F1s spectra of fluorinated BLG. (b) Raman spectra of non-fluorinated/pristine (BLG) and fluorinated bilayer graphene (F-BLG) membranes. (c) Absorption spectra of non-fluorinated (BLG) and fluorinated (F-BLG) bilayer graphene membranes.

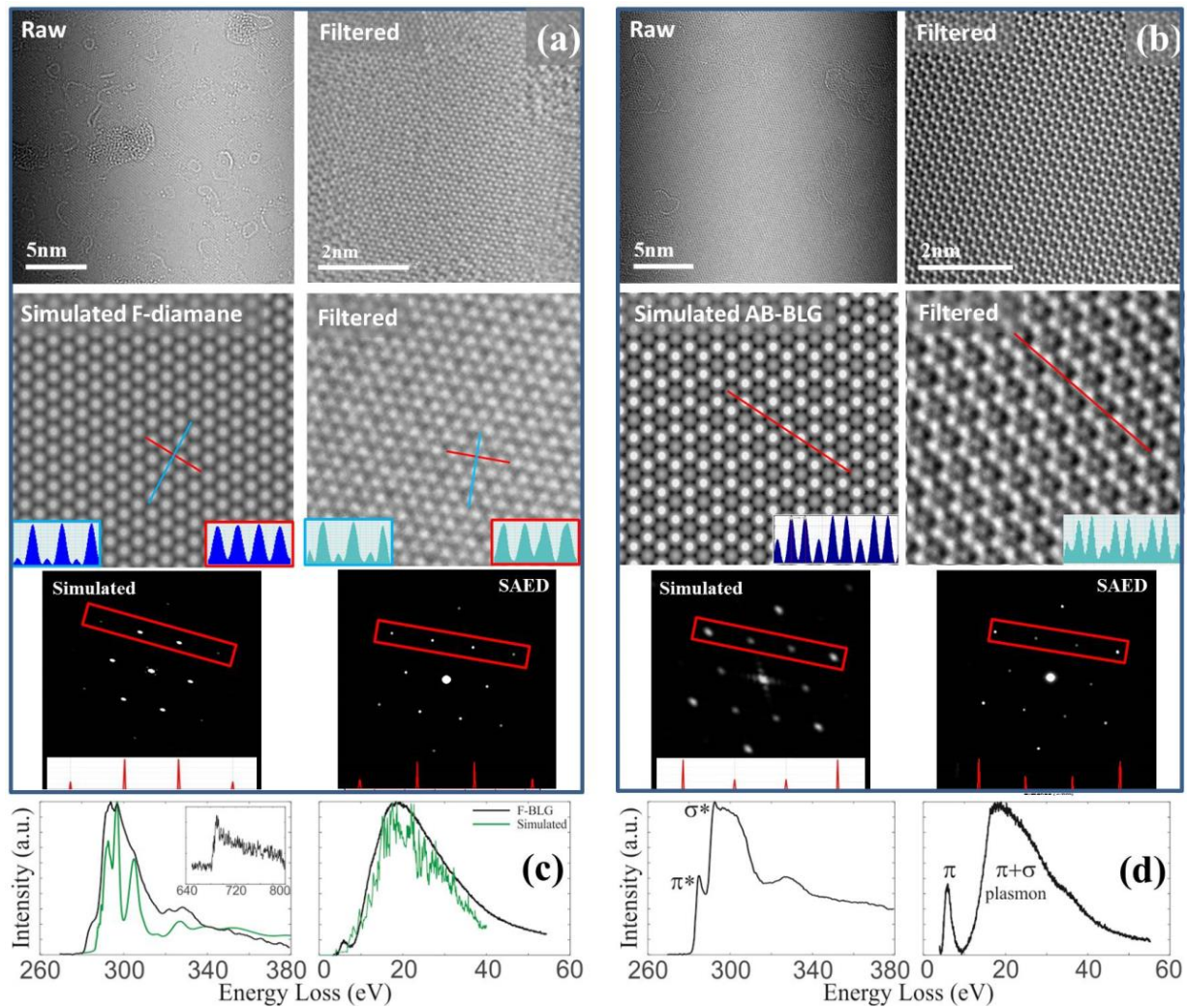

**Fig. 5. TEM and EELS studies of fluorinated bilayer graphene membranes.** Experimental and simulated HR-TEM images and SAED patterns obtained for (a) fluorinated BLG and (b) pristine AB-BLG. K-edge and low EEL spectra measured for (c) fluorinated BLG and (d) pristine AB-BLG.

# Supplementary Information for

## Chemically Induced Transformation of CVD-Grown Bilayer Graphene into Single Layer Diamond

Pavel V. Bakharev, Ming Huang, Manav Saxena, Suk Woo Lee, Se Hun Joo, Sung O Park, Jichen Dong, Dulce Camacho-Mojica, Sunghwan Jin, Youngwoo Kwon, Mandakini Biswal, Feng Ding, Sang Kyu Kwak, Zonghoon Lee & Rodney S. Ruoff

*correspondence to: rsruoff@ibs.re.kr, ruofflab@gmail.com (RSR)

bakharevpavel@gmail.com (PVB)

Density functional theory (DFT) Calculations:

*Formation energy calculation*:

To obtain optimized structures of fluorinated bilayer graphene (*i.e.*, two fluorinated graphene layers) which have the chemical formula of $C_2F$, density functional theory (DFT) calculations were performed using the CASTEP program (*40*). Generalized gradient approximation with Perdew-Burcke-Ernzerhof (GGA-PBE) functional (*41*) was used to describe the exchange-correlation potential of electrons. The electron-ion interaction was described by on-the-fly generated norm conserving pseudopotential. The van der Waals interaction was corrected by the Grimme's method (*42*). Spin polarization was taken into account in all the calculations. The electronic wave functions were expanded using a plane-wave basis set with the cutoff energy of 840 eV. The Brillouin-zone was sampled by the Monkhorst-Pack scheme (*43*) with the k-point separation of around 0.04 Å$^{-1}$. The self-consistent field (SCF) calculation was carried out with fixed orbital occupancy until the convergence criterion of $5 \times 10^7$ eV/atom was satisfied. The convergence criteria for the geometry optimization were set to $5 \times 10^{-6}$ eV/atom for the maximum energy change, 0.01 eV/Å for the maximum force, 0.02 GPa for the maximum stress, and 5 × 10⁻⁴ Å for the maximum displacement. Periodic boundary conditions were applied in x and y dimensions (*i.e.*, in-plane direction). The vacuum slab of 15 Å height was set along the z direction. For the optimized structures of the fluorinated bilayer graphene, the formation energy ($\Delta E_f$) was calculated as follows,

$$\Delta E_f = E_{C_2F} - 2E_C - \frac{1}{2}E_{F_2},$$

where $E_{C_2F}$, $E_C$, and $E_{F_2}$ are the total energy of fluorinated bilayer graphene per C₂F unit, the total energy of the bilayer graphene per C atom, and the total energy of F₂ molecule, respectively.

*Ionization potential calculations*:

The ionization potential (IP) of F-diamane was estimated by conducting spin-polarized DFT calculations using the DMol³ program (*44*). The exchange-correlation potential of electron was described by the GGA-PBE functional *(41)*. The core electrons were treated as all the electrons with relativistic effect. The DNP 4.4 basis set was used with a global orbital cutoff of 3.7 Å. The van der Waals interaction was corrected using the Grimme's method *(42)*. The convergence criterion of SCF calculation was set to 1.0 × 10⁻⁶. The smearing value of 0.005 Ha was applied. The convergence criteria for geometry optimization were set to 1.0 × 10⁻⁵ Ha for the maximum energy change, 0.002 Ha/Å for the maximum force, and 0.005 Å for the maximum displacement. The IPs of six F-diamane domains of a different size (Fig. S13) were calculated as

$$IP = E(M^+) - E(M),$$

where $E(M^+)$ and $E(M)$ are the energies of optimized cationic and neutral structures, respectively.

*Electronic and optical band gaps of pristine (defect-free) and 'defective' F-diamanes*:
The electronic and optical band gaps were calculated by applying the many-body electron-electron correlation theory using Green's function ($G_0W_0$) approximations and the electron-hole interaction (*i.e.*, excitonic) effects simulated by solving the Bethe-Salpeter equation (*BSE*) (*45-50*). These state-of-the-art methods are implemented in the Vienna *ab initio* simulation package (VASP) (*51,52*). The quasiparticle energies for the electronic band gaps of pristine and defective (with an F monovacancy per unit cell) F-diamanes were calculated using the wave functions of the model structures (Fig.S18) optimized by the GGA-PBE functional *(41)*. The electronic band gaps of 7.22 eV and 5.92 eV were obtained using the $G_0W_0$ approach for pristine and defective structures, respectively. The excitonic effects were then simulated by solving the *BSE* of the two-particle Green's function. The exciton binding energies of 1.38 eV and 3.05 eV were correspondingly calculated for pristine and defective F-diamanes leading to respective optical gaps of 5.84 eV for pristine F-diamane and 2.87 eV for the *defective* F-diamane structure. A strong effect of the F monovacancy defects on the electronic and optical band gaps can be rationalized by a relatively high defect density of 12.5% (a monovacancy per unit cell) resulting in $C_2F_{0.88}$ configuration of the modeled *defective* F-diamane structure.

For the DFT calculations with the *GGA-PBE* functional, the kinetic energy cutoff was set to 400 eV. The electron-ion interaction was described by the projector augmented-wave (PAW) method (*53*). The van der Waals interaction was corrected using the Grimme's method (*42*). The convergence criterion of electronic minimization was set to $1.0 \times 10^{-8}$ eV. The geometry optimization was performed for the maximum force of 0.01 eV/Å. The energy cutoff of 270 eV was used for $G_0W_0$. The *BSE* was solved using 8 occupied orbitals and 8 virtual orbitals. Spin-polarization was taken into account in all the calculations. The $11 \times 11 \times 1$ and $3 \times 3 \times 1$ gamma centered *k*-point meshes were used for pristine F-diamane and F-diamane with F monovacancy.

TEM image simulation:

HR-TEM image simulations were performed in two ways, namely by using the SimulaTEM code *(54)* and by multi-slice method using the MacTempas software.

EELS calculation:

The core electron energy loss spectroscopy calculation was carried out by using the CASTEP module as implemented in the Materials Studio (*55,56*). A 1×1 hexagonal diamond (111) slab composed of two carbon atomic layers with its two surfaces passivated with F was used to model fluorinated diamond layer structures. For the calculation, an energy cutoff of 760 eV was used for the plane wave basis set, and the interaction between valence electrons and ion cores was described by the OTFG norm conserving pseudopotential. The k-point mesh was sampled by 15×15×1. Other parameters are similar to those described above.

The electron energy loss spectroscopy in the low energy range was obtained through the same method as in the reference (*57*). The models used were the same as those used for core electron energy loss spectroscopy calculation. The calculation was performed by using the VASP code. Similar parameters to those in Fig.S7 were used, except for the fact that the k-point mesh was sampled by 17×17×1 and the energy was converged to be within $10^{-6}$ eV.

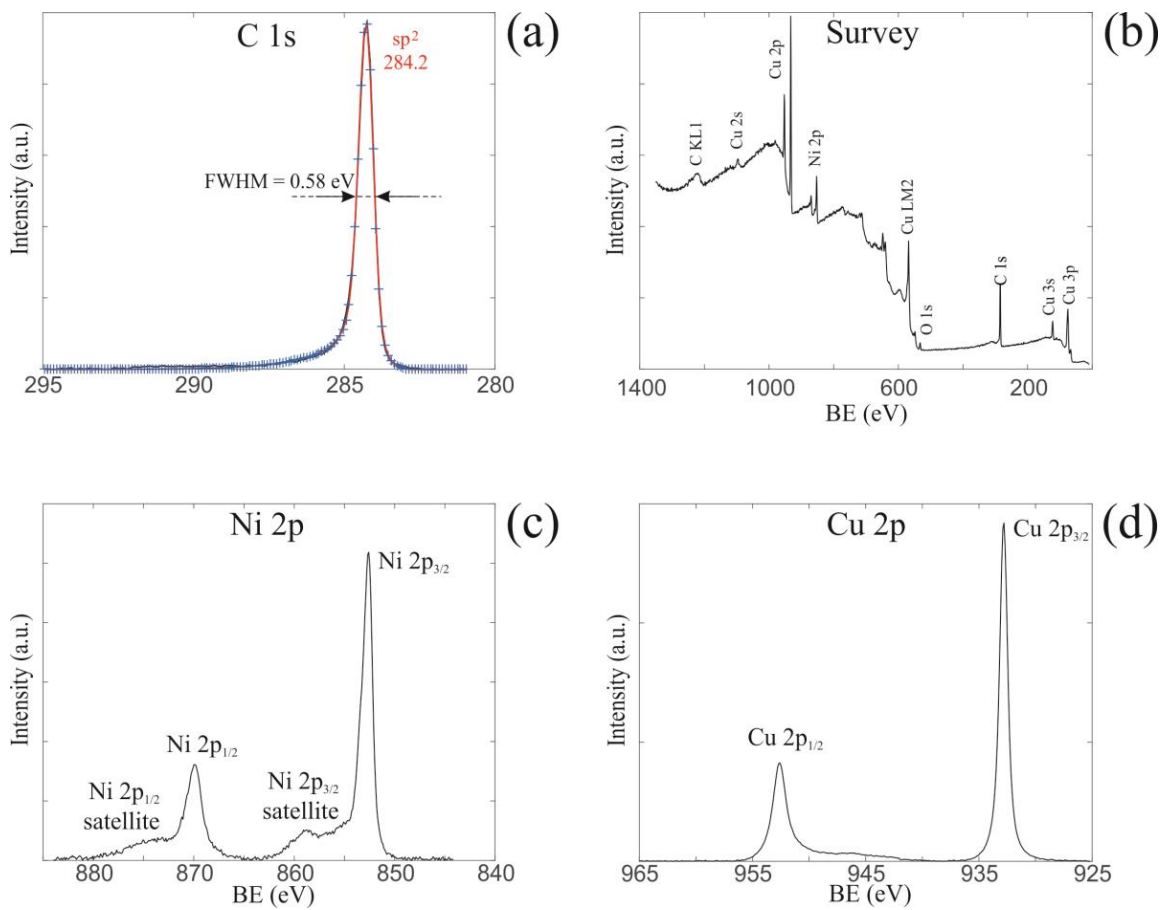

**Fig. S1**

XPS characterization of as-grown BLG on single crystal CuNi (111). (a) C1s, (b) Survey (c) Ni2p, and (d) Cu2p spectra.

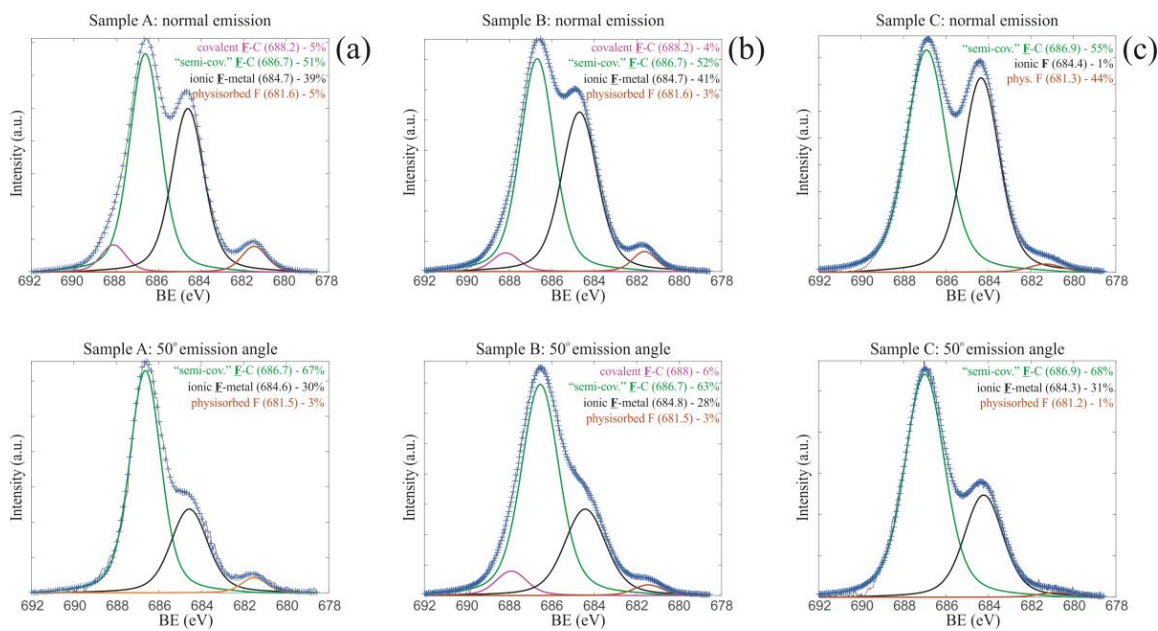

**Fig. S2**

F1s spectra obtained at 0° and 50° grazing emission angles for (a) Sample A (>12 hour fluorination), (b) Sample B (~6 hour fluorination) and (c) Sample C (2-3 hour fluorination).

**Table S1**. Fluorine content in Samples A, B and C calculated using C1s spectra deconvolution (column "C1s") and integrated intensity ratios of C1s spectra to corresponding primary "semi-covalent" F-C F1s peak (column "C1s/F1s").

|  | C1s | C1s/F1s |
|---|---|---|
| Sample C | 16 at.% | 15 at.% |
| Sample B | 26-27 at.% | 25 at.% |
| Sample A | 32 at.% | 32 at.% |

The content of fluorine covalently bonded to carbon was calculated using two approaches, namely, by deconvoluting XPS C1s spectra of fluorinated BLG samples and by analyzing the integrated intensity ratios, $I^{C1s}/I^{F1s}_{cov}$, of C1s spectra to the "semi-covalent" signal in the corresponding F1s spectra.

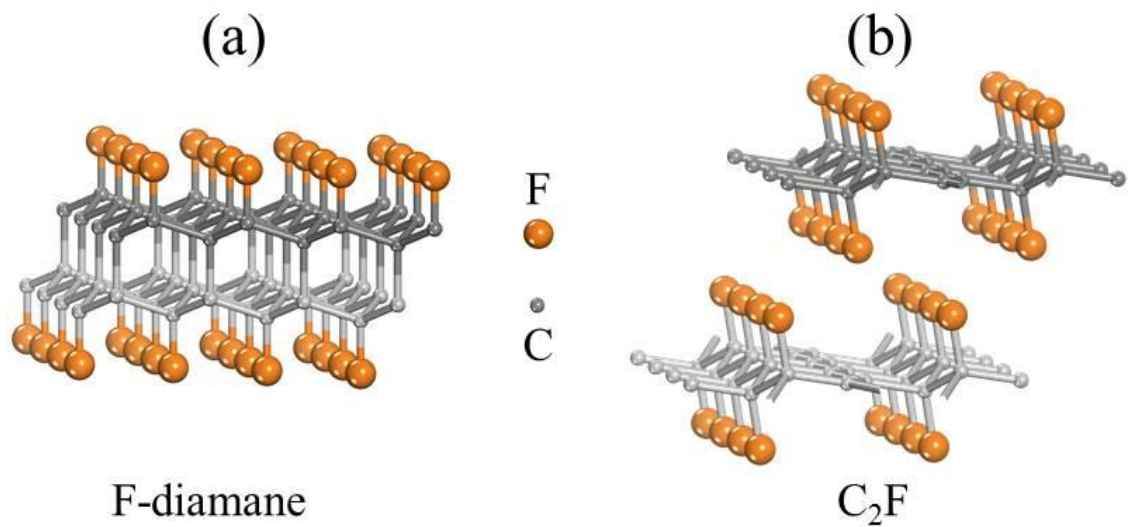

**Fig. S3**

Ball-and-stick models of (a) F-diamane and (b) the $C_2F$ structure without interlayer C-C bonds.

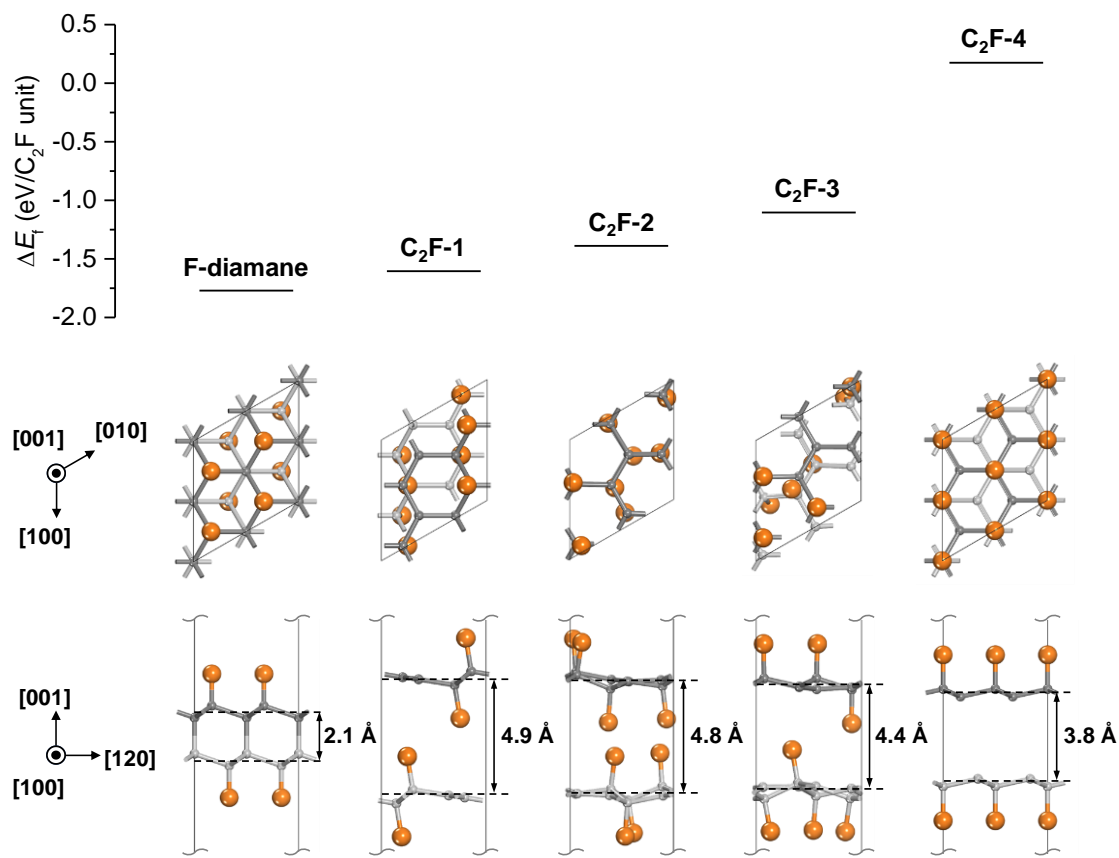

**Fig. S4**.

Formation energies ($\Delta E_f$) and [001] and [100] projection views of F-diamane and various $C_2F$ configurations with no interlayer bonds ($C_2F$-1, $C_2F$-2, $C_2F$-3, and $C_2F$-4). Orange, dark gray, and light gray represent fluorine, carbon in the top layer, and carbon in the bottom layer, respectively. The interlayer distance was calculated on the basis of the average z-coordinate of carbon atoms in each layer.

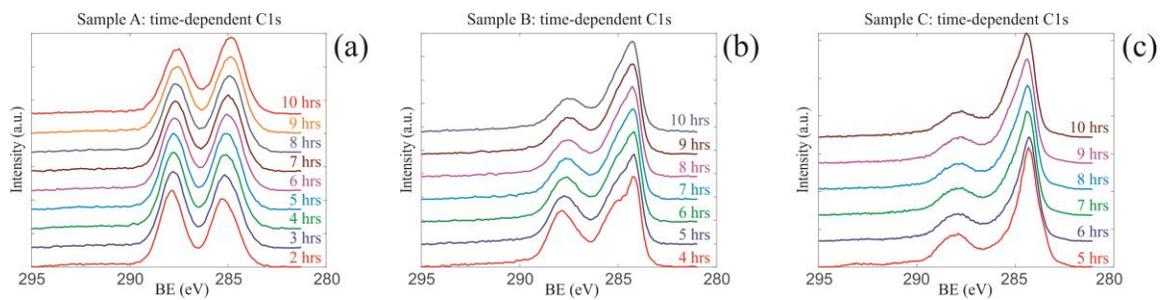

**Fig. S5**

Time-dependent XPS analysis performed for (a) Sample A, (b) Sample B and (c) Sample C.

**Table S2**. The average thickness of fluorinated BLG overlayers on the metal fluoride surface was calculated by measuring the intensity ratios $R^{C1s}_{F1s\_ionic}$ (the thickness values in the 2nd column), $R^{F1s\_cov}_{F1s\_ionic}$ (the thickness values in the 3rd column), $R^{C1s}_{F1s\_phys}$ (the thickness values in the 4th column) and $R^{F1s\_cov}_{F1s\_phys}$ (the thickness values in the 5th column). The error is ±0.07 nm.

|  | $R^{C1s}_{F1s\_ionic}$, nm | $R^{F1s\_cov}_{F1s\_ionic}$, nm | $R^{C1s}_{F1s\_phys}$, nm | $R^{F1s\_cov}_{F1s\_phys}$, nm |
|---|---|---|---|---|
| Sample C | 0.63 | 0.66 | weak signal from physisorbed F | weak signal from physisorbed F |
| Sample B | 0.62 | 0.63 | 0.55 | 0.57 |
| Sample A | 0.60 | 0.56 | 0.61 | 0.57 |

We used AR-XPS and the uniform overlayer model to calculate the average thickness, $d$, of fluorinated BLG films (*58*). According to this model, the average thickness, $d$, can be determined from the angular dependence of the intensity ratio, $R^{ovl}_{sub}$, of overlayer and substrate signals. Thus, the average thickness, $d_{CF}$, of fluorinated BLG on the metal fluoride surface was estimated in four ways, namely 1) from the integrated intensity ratios, $R^{C1s}_{F1s\_ionic}$, of the C1s spectra to the ionic (metal fluoride) signals in the corresponding F1s spectra (the thickness values in the 2nd column in Table S2); 2) from the integrated intensity ratios, $R^{F1s\_cov}_{F1s\_ionic}$, of the covalent signal (C-F) to the ionic signal in the F1s spectra (the thickness values in the 3rd column in Table S2); 3) from the integrated intensity ratios, $R^{C1s}_{F1s\_phys}$, of the C1s spectra to the physisorbed/entrapped signal in the F1s spectra (the thickness values in the 4th column in Table S2); 4) from the integrated intensity ratios, $R^{F1s\_cov}_{F1s\_phys}$, of the covalent signal (C-F) to the physisorbed/entrapped signal in the F1s spectra (the thickness values in the 5th column in Table S2), which are given by

$$R^{C1s}_{F1s_{ionic}}(\theta) \equiv \frac{I^{C1s}}{I^{F1s}_{ionic}} = \frac{\sigma_C(h\nu)\lambda(E^{C1s}_{kin})T(E^{C1s}_{kin})N_C}{\sigma_F(h\nu)\lambda(E^{F1s}_{kin})T(E^{F1s}_{kin})N^{ionic}_F}\left(e^{\frac{d_{CF}}{\lambda(E^{F1s}_{kin})\cos\theta}} - 1\right);$$

$$R^{F1s\_cov}_{F1s\_ionic}(\theta) \equiv \frac{I^{F1s}_{cov}}{I^{F1s}_{ionic}} \cong \frac{N^{cov}_F}{N^{ionic}_F}\left(e^{\frac{d_{CF}}{\lambda(E^{F1s}_{kin})\cos\theta}} - 1\right);$$

$$R^{C1s}_{F1s_{phys}}(\theta) \equiv \frac{I^{C1s}}{I^{F1s}_{phys}} = \frac{\sigma_C(h\nu)\lambda(E^{C1s}_{kin})T(E^{C1s}_{kin})N_C}{\sigma_F(h\nu)\lambda(E^{F1s}_{kin})T(E^{F1s}_{kin})N^{phys}_F}\left(e^{\frac{d_{CF}}{\lambda(E^{F1s}_{kin})\cos\theta}} - 1\right);$$

$$R^{F1s\_cov}_{F1s\_phys}(\theta) \equiv \frac{I^{F1s}_{cov}}{I^{F1s}_{phys}} \cong \frac{N^{cov}_F}{N^{phys}_F}\left(e^{\frac{d_{CF}}{\lambda(E^{F1s}_{kin})\cos\theta}} - 1\right),$$

where

σ (*hv*) is the photon absorption cross section;

λ ($E_{kin}$) is the inelastic mean free path of photoemitted electrons at kinetic energy $E_{kin}$;

$T(E_{kin})$ is the transmission function of the analyzer;

$N$ is the atom density;

$\theta$ is the angle between the surface normal and the direction toward the detector (emission angle).

The fractions outside of the parenthesis can be eliminated by taking the ratio of the intensity ratios measured at 0° and 50° emission angles.

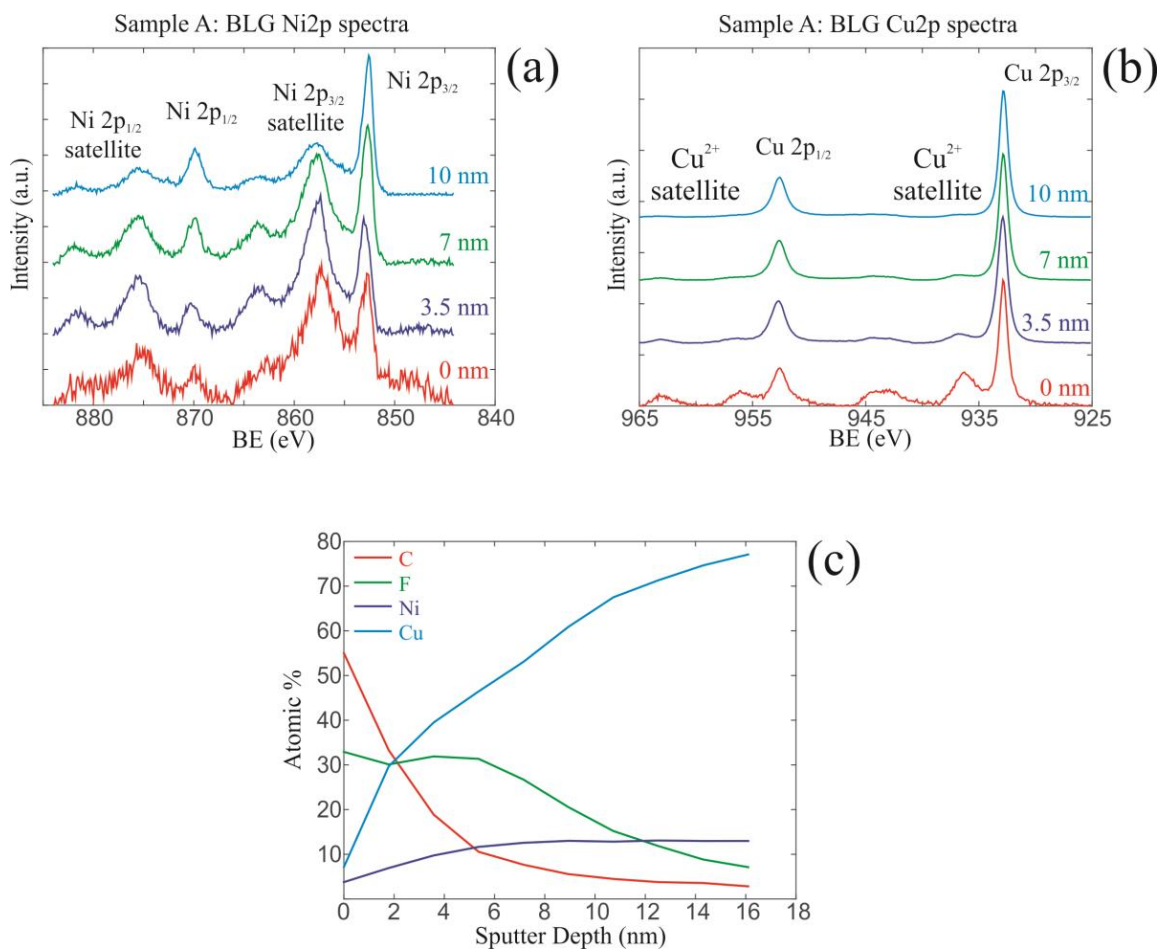

**Fig. S6**

XPS depth profiling of Sample A (Ni content ~16 at%). High resolution (a) Ni2p, (b) Cu2p spectra and (c) relative content of C, F, Ni and Cu as functions of depth.

In order to study the distribution of fluorine in fluorinated samples as a function of depth, we conducted an XPS depth profiling analysis by using Ar+ ion-beam at 3 keV to etch/sputter the surface. As it can be inferred from the data provided in Fig. S6, fluorine and metal fluoride signals were detected not only on the surface but also in the bulk of the substrate down to at least 10 nm depth indicating fluorine diffusion into the metal alloy.

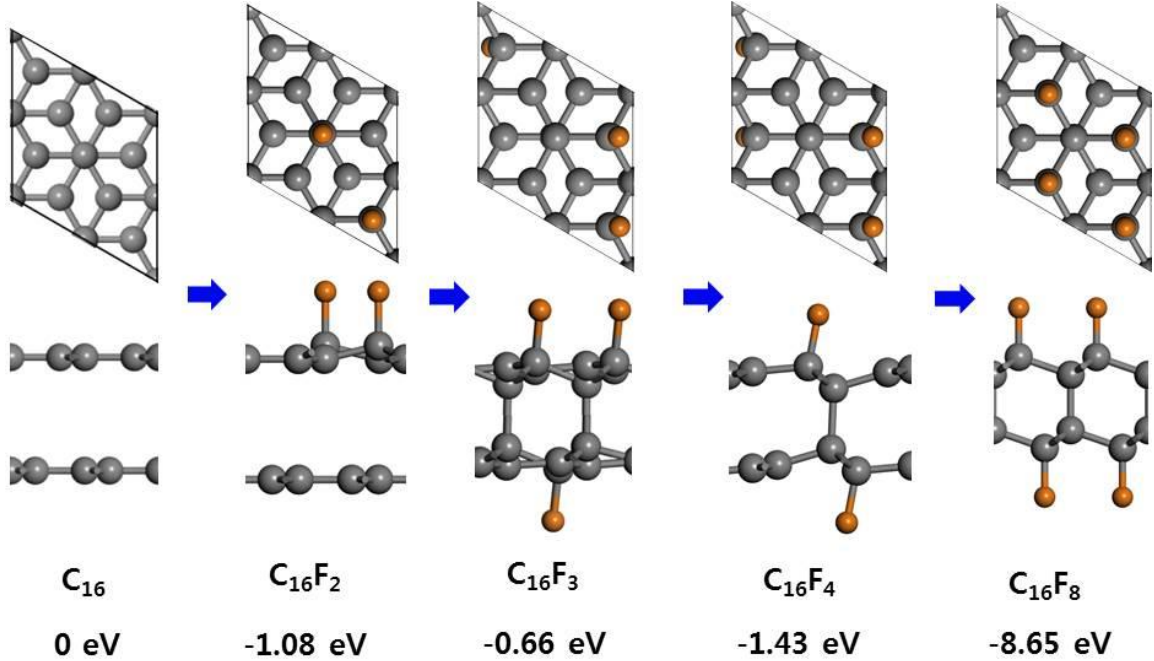

**Fig.S7**

Optimized structures of pristine BLG ($C_{16}$) and fluorinated BLG of different stoichiometry, namely $C_{16}F_2$, $C_{16}F_3$, $C_{16}F_4$, $C_{16}F_8$ (F-diamane) with corresponding relative energies.

To compare relative stabilities of different structures, we chose pristine bilayer graphene and an F atom in a $XeF_2$ molecule as the reference. The relative energies of different structures were calculated by the following equation:

$$E_{Rel} = E_i - E_0 - n_i \times \varepsilon_F,$$

where $E_i$ is the total energy of the $i^{th}$ structure during the fluorination process, $E_0$ is the energy of pristine bilayer graphene, $n_i$ is the number of F atoms in the $i^{th}$ structure and $\varepsilon_F$ is the energy of a F atom in $XeF_2$ molecule, estimated by

$$\varepsilon_F = (\varepsilon_{X_eF_2} - \varepsilon_{X_e})/2,$$

where $\varepsilon_{X_eF_2}$ is the energy of a $XeF_2$ molecule and $\varepsilon_{X_e}$ is the energy of an Xe atom.

Pristine bilayer graphene is modeled by a 2×2 unit cell with a vacuum layer of 15 Å. During fluorination, F atoms are added to the bilayer graphene structure. The structures were optimized through the conjugated gradient method until the forces on each atom were less than 0.01 eV/Å, with an energy convergent criteria of $10^{-4}$ eV. The optimization was carried out by using the first-principles density functional theory (DFT) calculations as implemented in the Vienna *ab* *initio* simulation Package (VASP) code (*59,52*). To take the weak van der Waals interaction into consideration, the DFT-D2 method was adopted (*42*). The exchange-correlation effect was treated by the Perdew-Burke-Ernzerhof generalized gradient approximation (GGA) (*41*). The projected augmented wave (PAW) method was employed to describe the interactions between valence electrons and ion cores (*53*). The k-point mesh was sampled by 9×9×1.

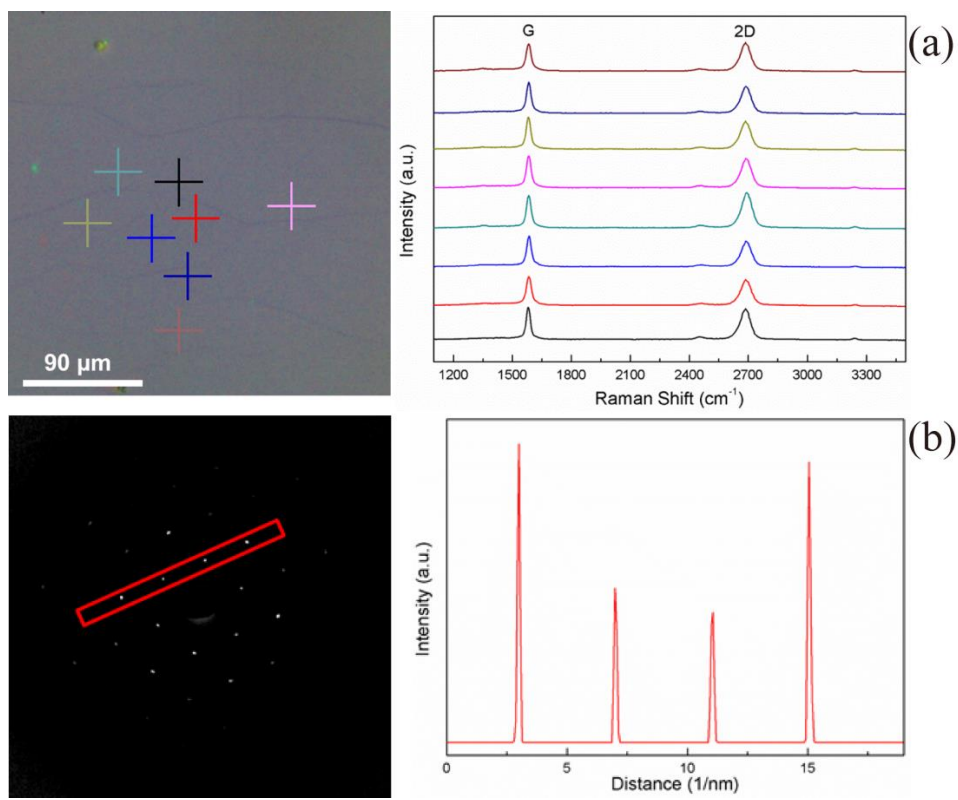

**Fig. S8**

(a) Optical micrograph and Raman spectra at 8 randomly chosen positions by 532 nm excitation of AB-stacked bilayer graphene (AB-BLG) grown by CVD on single crystal CuNi(111) and transferred onto a $SiO_2$/Si wafer (thickness of $SiO_2$ layer is 300 nm). (b) Selected area electron diffraction (SAED) pattern of AB-BLG and profile plots of the diffraction peak intensities obtained for the SAED pattern region highlighted in red.

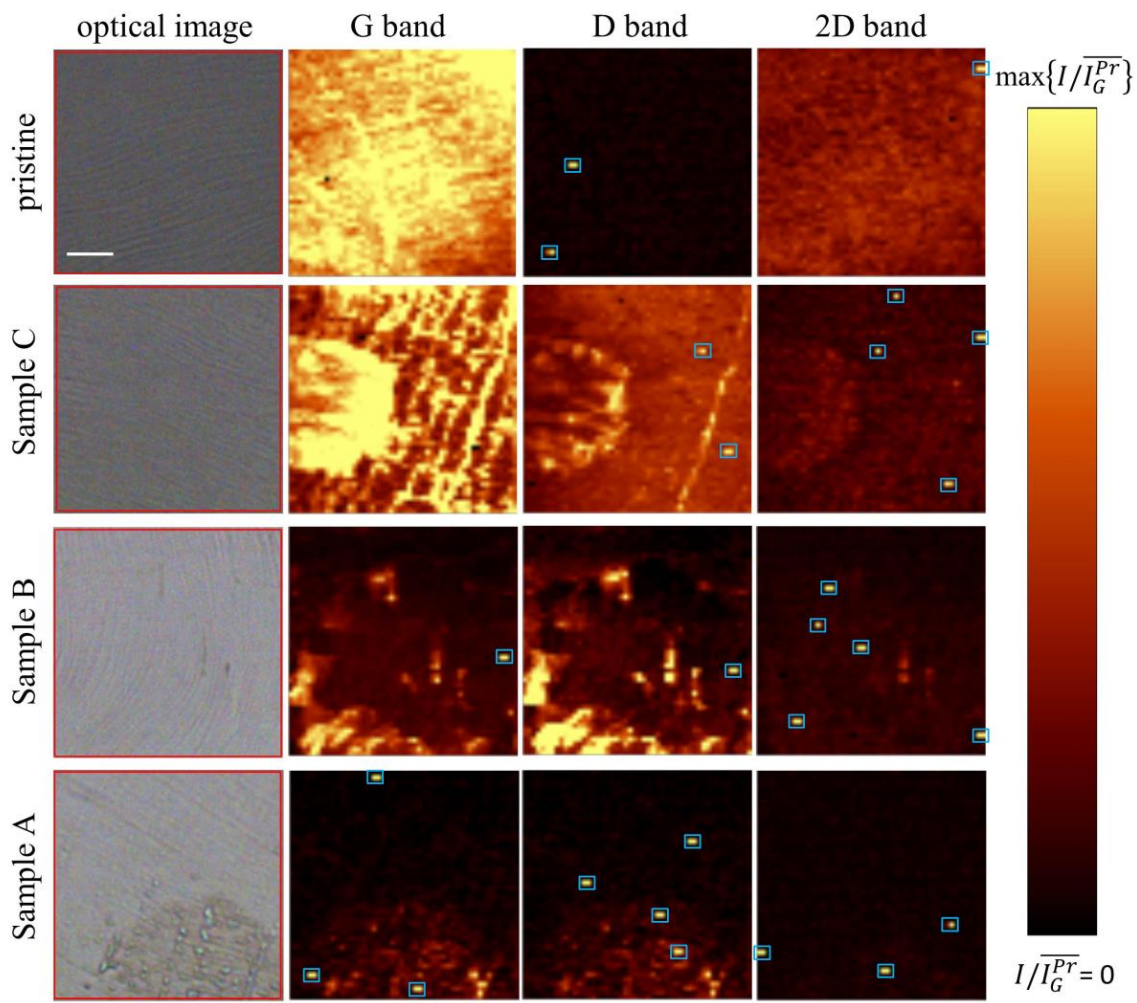

**Fig. S9**

Optical microscope images and Raman intensity ratio maps obtained from 25×25 μm² areas on as-grown BLG and Samples A, B, C. Ratios of G, D and 2D peak intensities to the average G peak maximum intensity of as-grown graphene. The intense spikes triggered by cosmic ray noise are highlighted with light blue rectangles. The scale bar is 5 μm.

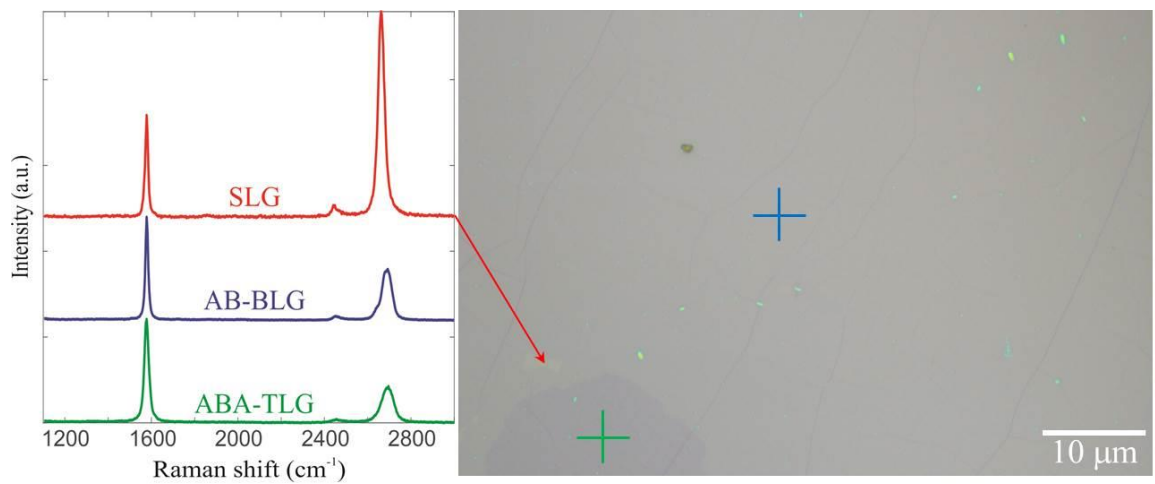

**Fig. S10.**

Optical and Raman characterization of as-grown graphene film transferred onto SiO$_2$/Si wafer (with 300 nm thick SiO$_2$ layer on the surface). The representative Raman spectra of continuous AB-stacked BLG, small region of SLG and ABA-stacked TLG domain of hexagonal shape.

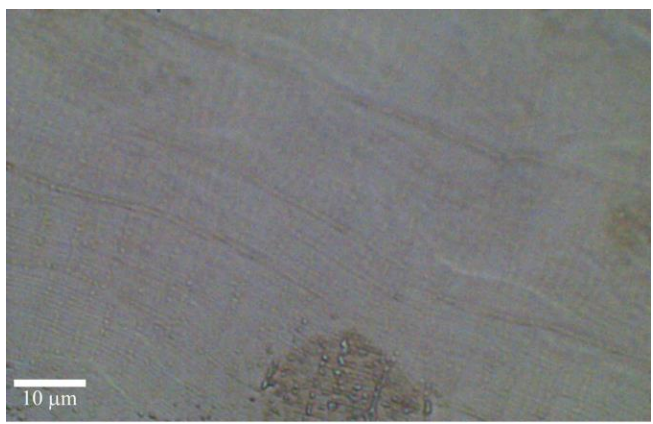

**Fig. S11**

Low magnification optical micrograph of Sample A. The image was taken from the region with multilayer graphene (MLG) domain.

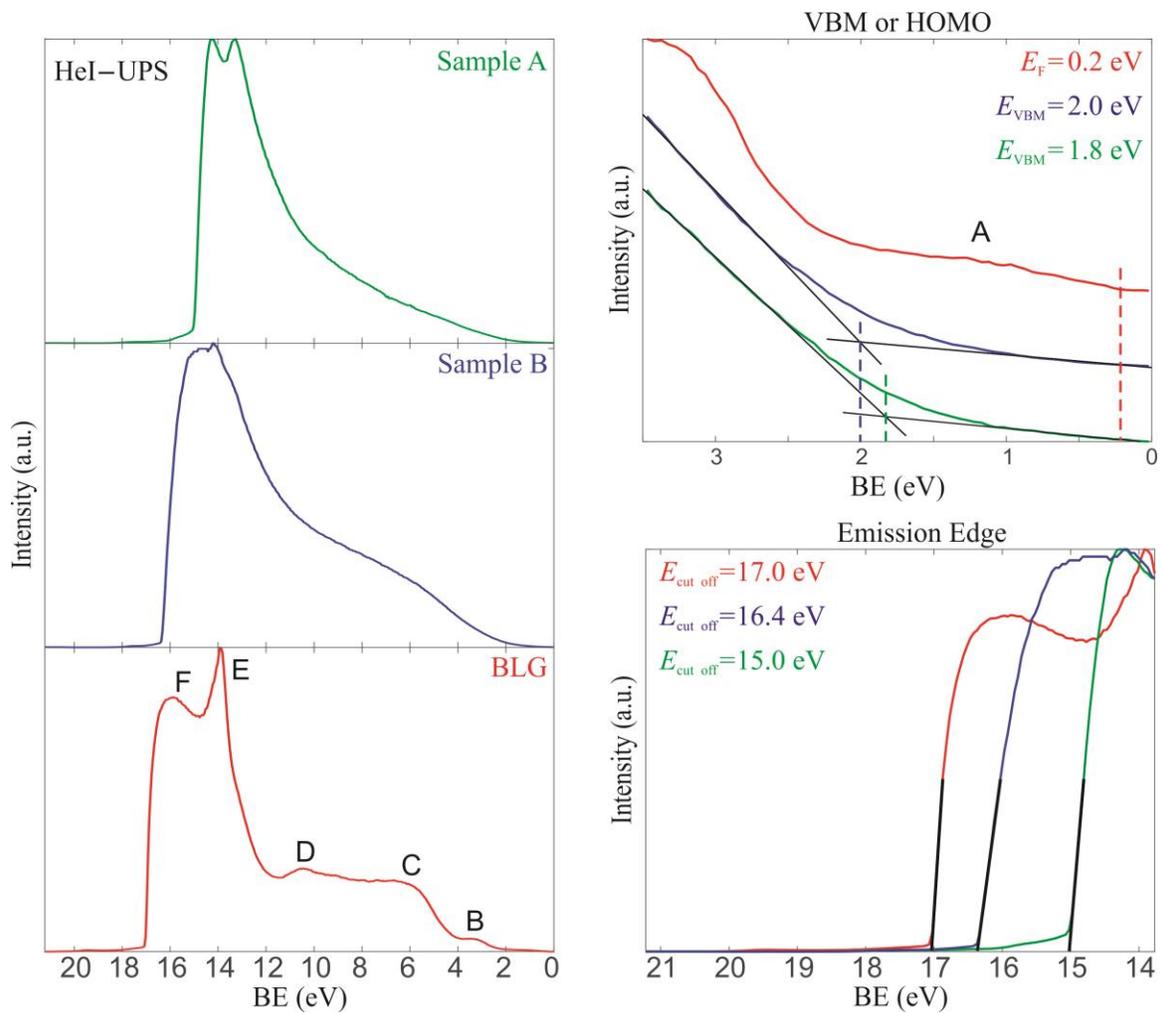

**Fig. S12**

UPS spectra acquired at 20º emission angle of as-grown BLG on the CuNi(111) surface, Sample A and Sample B.

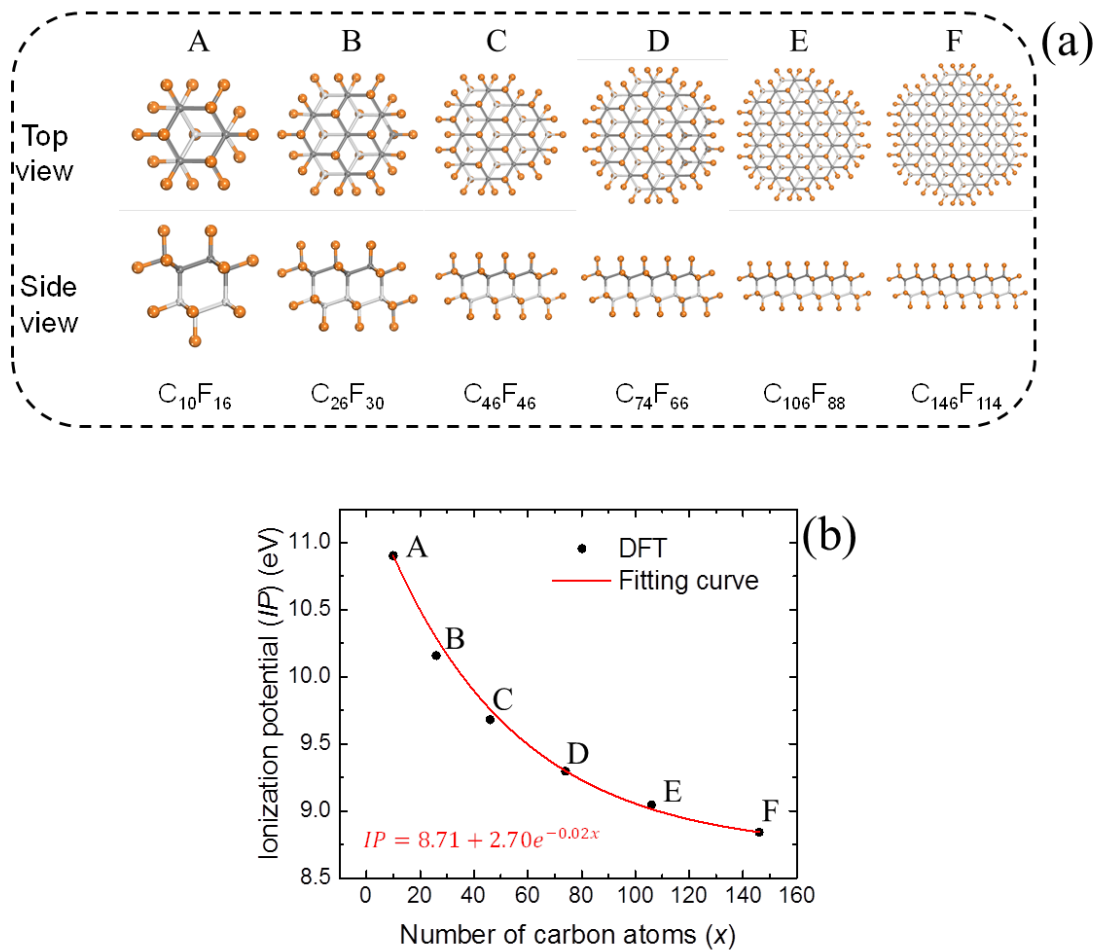

**Fig. S13**

(a) Top and side views of the optimized structures of six F-diamane domains of different size with (b) the corresponding calculated ionization potentials (black dotes) and the fitted curve (red curve).

The ionization potential (IP) of F-diamane was estimated by fitting/extrapolating the calculated IP values obtained for the F-diamane domains of different sizes. The estimated IP value of ~8.71 eV is in a good agreement with the experimental value of ~8 eV measured by UPS.

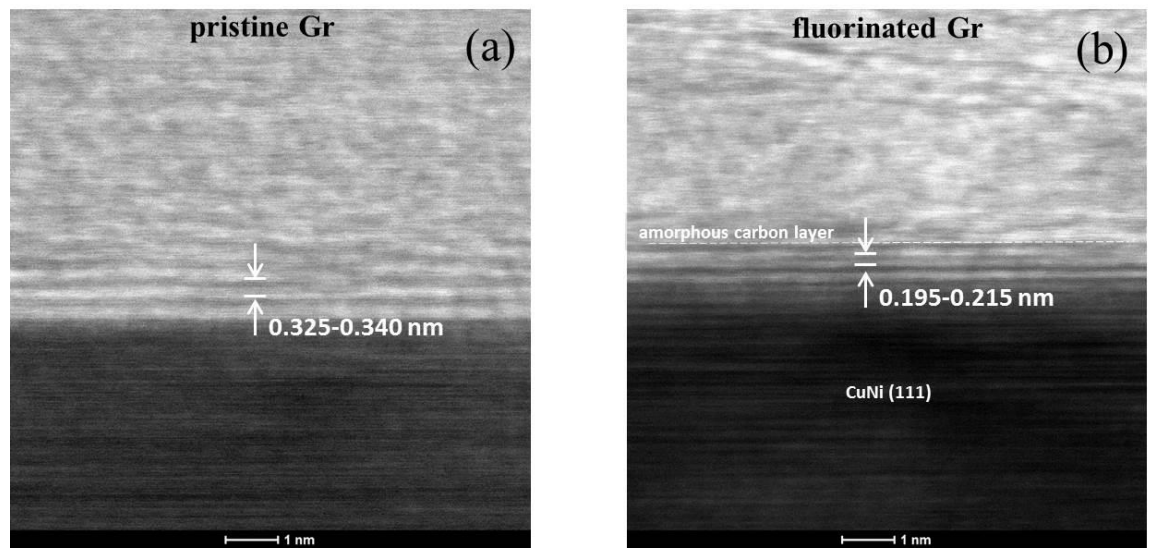

**Fig. S14**

Scanning transmission electron (STEM) micrographs (a) pristine BLG and (b) fluorinated BLG (Sample A) on single crystal CuNi(111).

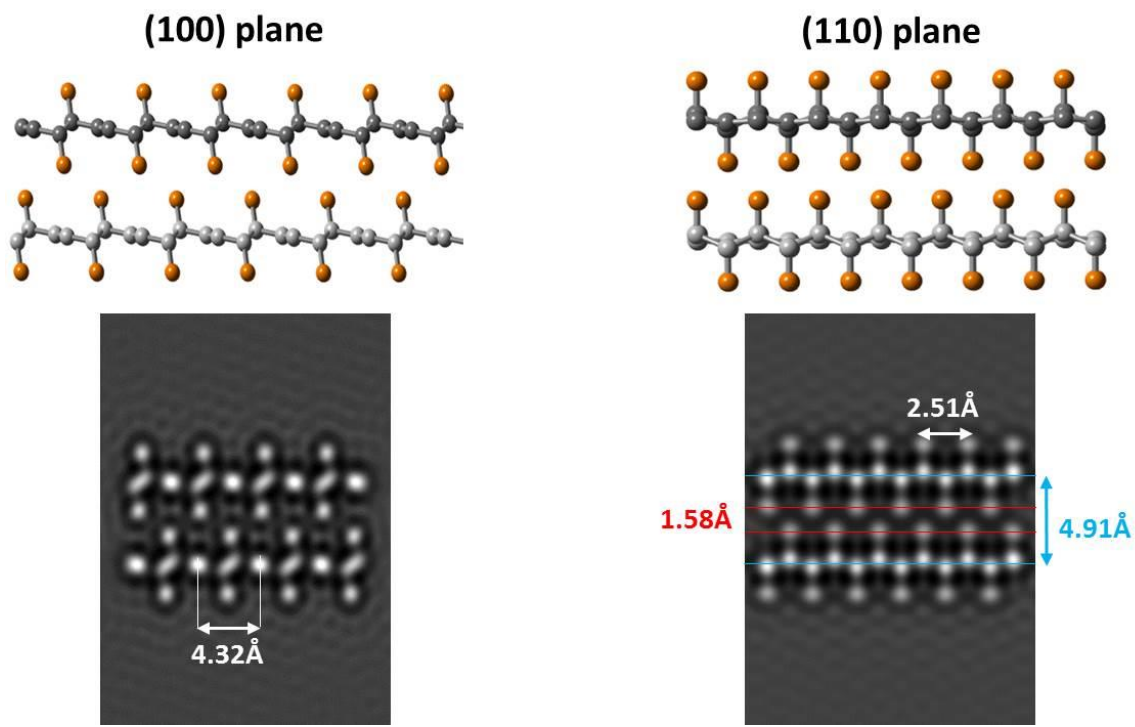

**Fig.S15**

Atomic models and simulated HR-TEM images of the DFT-optimized $C_2F$ structure without interlayer C-C linkages.

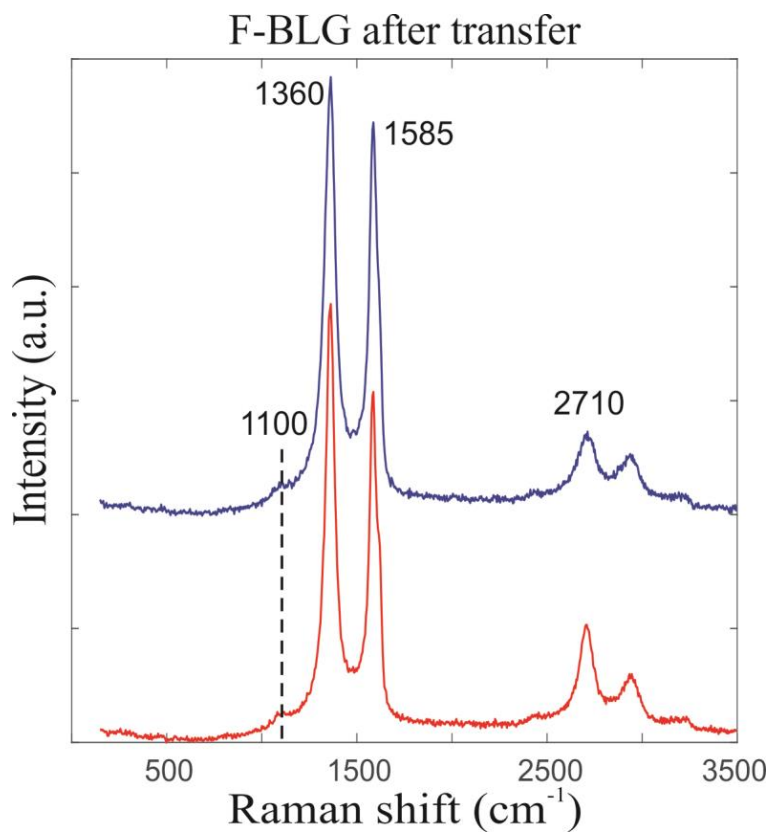

**Fig. S16**

Raman spectra at two randomly chosen positions by 488 nm excitation of suspended fluorinated BLG (F-BLG) measured after its transfer from the metal surface onto a Au TEM grid using electrochemical bubbling delamination technique.

The positions of the D at 1360 cm$^{-1}$, G at 1585 cm$^{-1}$ and 2D/G' at 2710 cm$^{-1}$ peaks are typical of those of chemically modified graphene films. However, the presence of sharp D and G peaks as well as the recovery of a prominent narrow 2D/G' band suggest that the sample undergoes extensive defluorination during the transfer process. A small peak at ~1100 cm$^{-1}$ might possibly be ascribed to the previously reported C-C sp$^3$ vibrations in carbon films *(60-62)*.

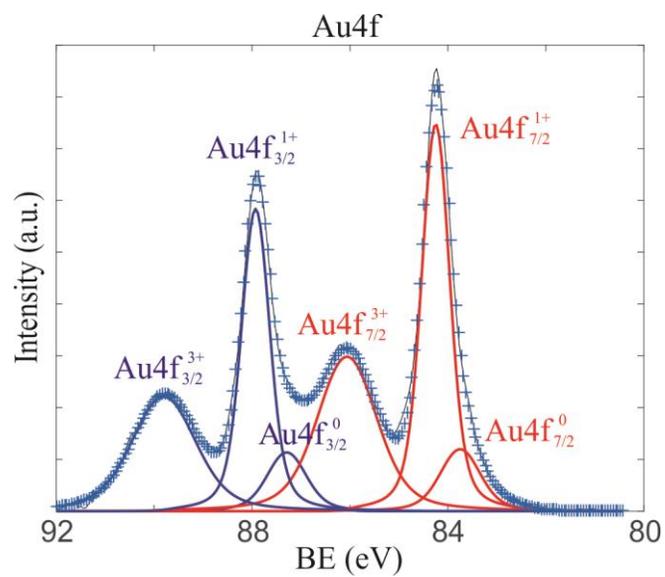

**Fig. S17**

High resolution Au4f spectra of a TEM grid after exposure to XeF$_2$ vapor for 6 hours at 45°C.

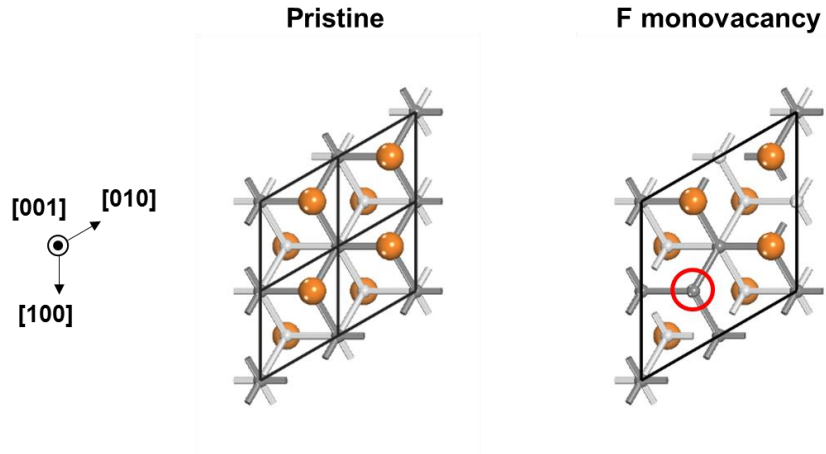

**Fig.S18**

The optimized structures of pristine (defect-free) F-diamane and F-diamane with F *monovacancy*. The F monovacancy site is highlighted with a red circle. The density of monovacancies is 12.5% which corresponds to $C_2F_{0.88}$ configuration of '*defective*' F-diamane.

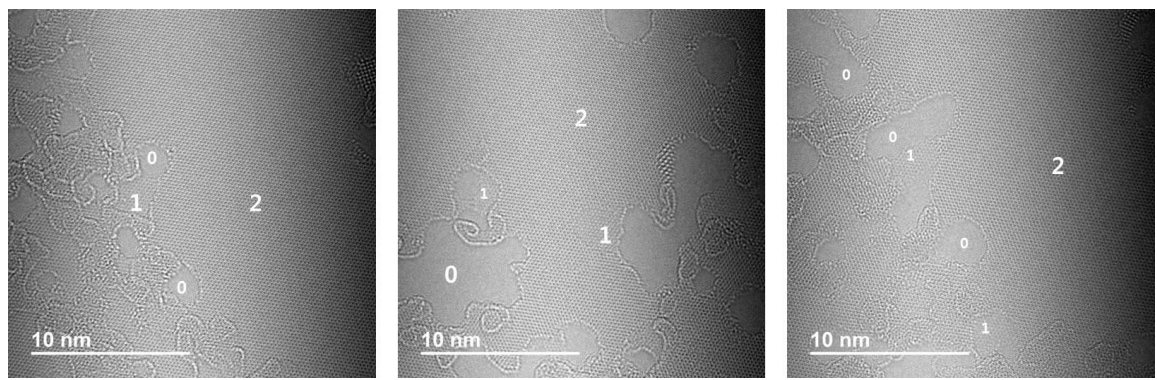

**Fig. S19**

HR-TEM images of graphene films after prolonged (more than 2 min) exposure to the electron beam. Regions containing holes, a graphene monolayer and a bi-layer are highlighted with numbers 0, 1 and 2, respectively.

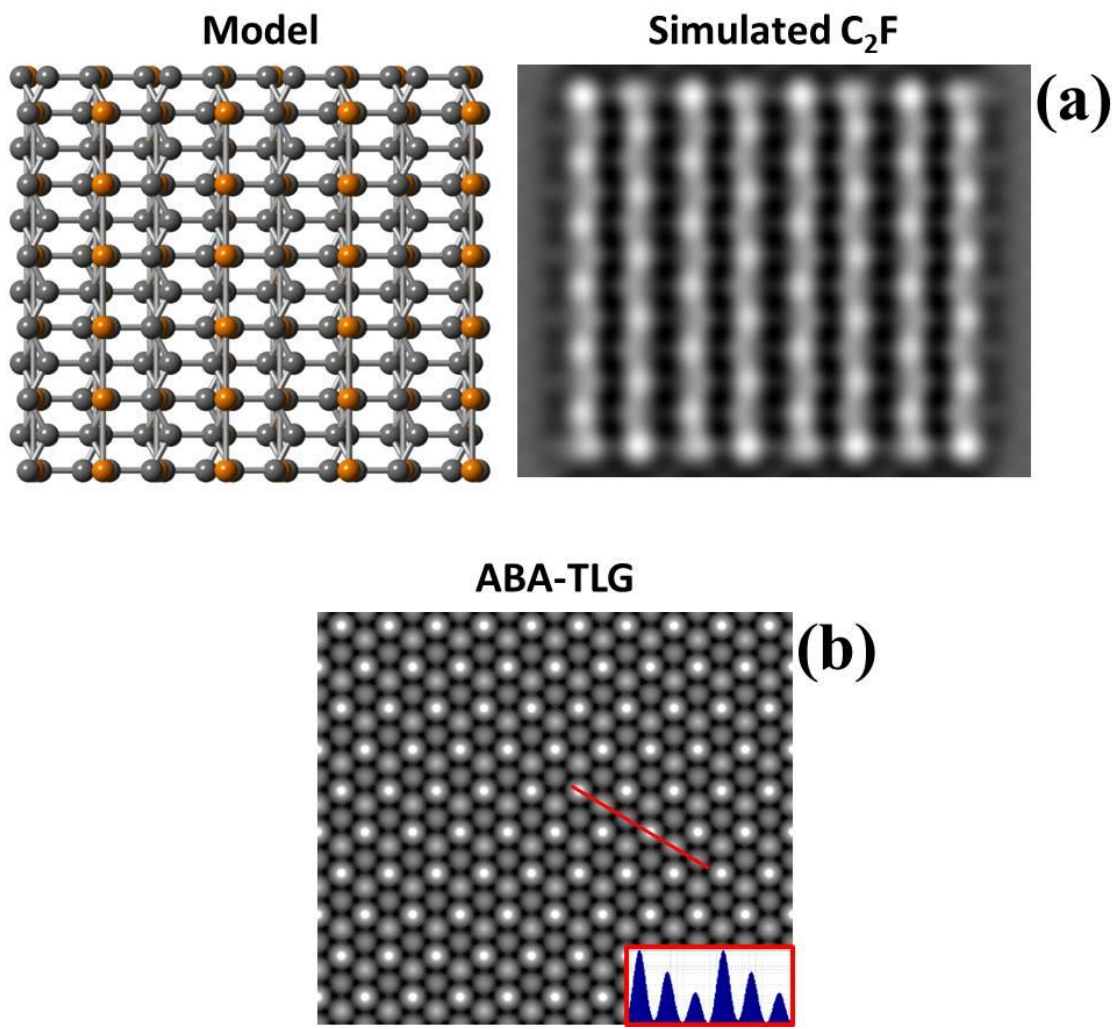

**Fig. S20**

Simulated HR-TEM images of the DFT-optimized (a) $C_2F$ structure without C-C interlayer bonds and (b) ABA stacked trilayer graphene.

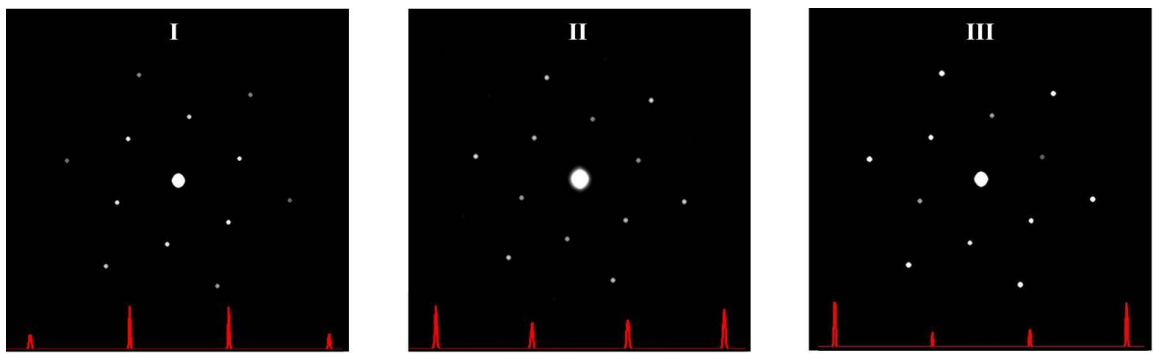

**Fig. S21**

Effect of electron dose rate on SAED pattern of fluorinated BLG.

The diffraction pattern in stage I (F-diamane) was acquired using a parallel electron beam of 11.1 μm size. Stages II (intermediate) and III (AB-BLG) correspond to the exposure to the parallel electron beam of 0.9 μm size for 20-30 sec and for >30 sec, respectively. The dose rate in stages II and III is ~150 times higher than the dose rate in stage I. The aperture size (the characteristic size of the analyzed area) in stages I, II and III is ~0.7 μm. Note that the exposure to the electron beam with an increased dose rate for more than 2 min results in damage/etching of the graphene films (Fig.S19).

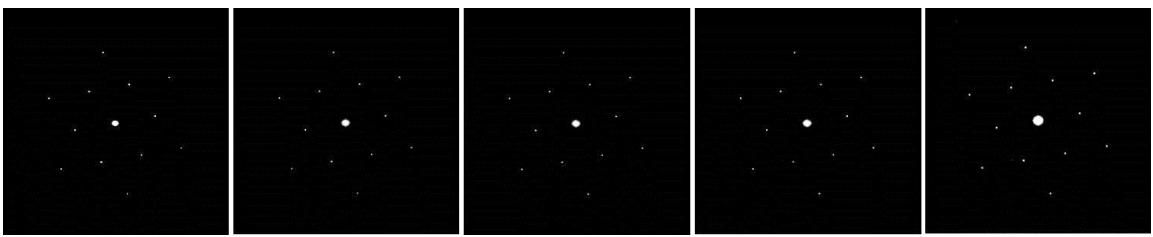

**Fig. S22**

SAED patterns collected at different locations in fluorinated BLG. Each SAED pattern was acquired from the circular region of ~0.7 μm diameter. The characteristic size of the analyzed regions and the alignment of the diffraction patterns acquired from different locations indicate a single crystal structure over relatively large area.